\documentclass[aps,superscriptaddress,amsmath,amssymb,twocolumn,amsfonts]{revtex4-2}
\usepackage{bm,xcolor,MnSymbol}
\usepackage{graphicx}
\graphicspath{{Pictures/}}
\usepackage{braket,natbib}  
\usepackage[colorlinks=true,linkcolor=magenta,urlcolor=purple,citecolor=magenta,anchorcolor=blue]{hyperref}
\usepackage[normalem]{ulem}
\usepackage{multirow}
\usepackage{orcidlink}

\newcommand{\dt}{\ensuremath{\mathcal{T}^\ast}}
\newcommand{\dpp}{\ensuremath{\mathcal{P}^\ast}}
\newcommand{\dss}{\ensuremath{\mathcal{S}}}
\newcommand{\doo}{\ensuremath{\mathcal{O}}}
\newcommand{\dr}{\ensuremath{\mathbf{r}}}
\newcommand{\vk}{\ensuremath{\mathbf{k}}}
\newcommand{\vq}{\ensuremath{\mathbf{q}}}
\newcommand{\vQ}{\ensuremath{\mathbf{Q}}}
\newcommand{\drr}{\ensuremath{\mathbf{R}}}
\newcommand{\df}{\ensuremath{\mathcal{F}}}

\newcommand{\bl}[1] {{\color{black}#1}}

\newcommand{\mb}[1]{\makebox[15pt]{#1}}

\begin{document}

\title{Eight-fold classification of superconducting orders}

\author{Alexander V. Balatsky\,\orcidlink{0000-0003-4984-889X}}
\email{alexander.balatsky@uconn.edu}
\affiliation{Nordita, KTH Royal Institute of Technology and Stockholm University, 106 91 Stockholm, Sweden}
\affiliation{Department of Physics, University of Connecticut, Storrs, Connecticut 06269, USA}
\author{Saikat Banerjee\,\orcidlink{0000-0002-3397-0308}}
\email{saikat.banerjee@uni-greifswald.de}
\affiliation{Institute of Physics, University of Greifswald, Felix-Hausdorff-Stra{\ss}e 6, 17489 Greifswald, Germany}
\affiliation{Center for Materials Theory, Rutgers University, Piscataway, New Jersey, 08854, USA}
\date{\today}

\begin{abstract}
\bl{We present a symmetry-based classification for superconducting pairing states, organized by the exchange properties of the anomalous correlation function rather than by a specific microscopic pairing mechanism. The classification is built from the pairwise permutation of spin, orbital, spatial, and temporal indices, leading to the fermionic constraint $\dss \dpp \doo \dt = -1$, and is further organized by separating relative and center-of-mass space-time coordinates. This construction defines what we call the Berezinskii--Abrahams hypercube, in which conventional Bardeen--Cooper--Schrieffer superconductivity, unconventional \(p\)- and \(d\)-wave pairing, odd-frequency superconductivity, Fulde--Ferrell--Larkin--Ovchinnikov states, pair-density-wave states, and time-modulated superconducting orders appear as different sectors of a unified framework. Beyond organizing known phases, the Berezinskii--Abrahams hypercube identifies symmetry-allowed hybrid orders that have received comparatively little attention, including odd-frequency Fulde--Ferrell--Larkin--Ovchinnikov or pair-density-wave states and odd-frequency time-modulated superconducting states. We discuss   microscopic routes, candidate platforms, experimental signatures, and stability constraints for these sectors, emphasizing the distinction between symmetry allowance and physical realizability. We also present the proximity induced odd-frequency pair-density-wave state, and its driven analog  as the two new examples of the states that naturally emerge in the Berezinskii--Abrahams hypercube. 
The resulting framework provides a guide for connecting established superconducting phenomena with unexplored symmetry-allowed forms of order.
}
\end{abstract}

\maketitle

\section{Introduction \label{sec:sec.1}}

Superconductivity is a remarkable example of a quantum state of matter that can occur on a 
macroscopic length scale. When a material becomes a superconductor, a significant fraction of the electrons near its Fermi surface condenses into a ``superfluid'' that extends over the entire system. This feature leads to macroscopic electronic flow with vanishingly small resistance. Since its discovery more than a century ago~\cite{Onnes1911}, superconductivity has remained at the center of condensed-matter research. To date, a wide range of superconducting materials and mechanisms have been discovered: conventional superconductors such as Al or Nb, which are well described by the seminal Bardeen--Cooper--Schrieffer (BCS) theory~\cite{PhysRev.108.1175}; high-$T_c$ cuprate superconductors, for which a complete microscopic description is still lacking~\cite{RevModPhys.92.031001}; organic superconductors~\cite{Jerome01081982,Ishiguro1990,Saito2011}; iron-based superconductors~\cite{Li_2011,PhysRevX.6.041045,Kreisel2020}; and topological superconducting systems. Recent interest in topology has led to a closer examination of superconductors constrained by crystalline, spin, and internal symmetries, particularly because of their potential relevance for realizing Majorana fermions~\cite{RevModPhys.83.1057}. \bl{This diversity of superconducting platforms motivates a classification principle that does not rely on a specific microscopic pairing mechanism, but instead organizes possible superconducting states according to their symmetry properties.} 

Superconductivity requires the development of long-range two-fermion correlation $\df_{\alpha\beta,ab}(\dr, \dr'\,|\,t,t') = \braket{\mathcal{T}_t c_{\alpha a}(\dr,t) c_{\beta b}(\dr',t')}$ that describes the pairing correlations in superconductors.  Here, $\mathcal{T}_t$ represents the time ordering operator (contour ordering in case of Keldysh analysis), and $\dr,\dr'$ and $t,t'$ are spatial and temporal coordinates of the two particles, respectively; $\{a,b\}$ denotes respective orbital or band degrees of freedom, while $\{\alpha,\beta\}$ are the spin degrees of freedom. \bl{In the following, we 
refer to \(\df\) as the anomalous Green's function, 
while reserving \(\Delta\) for the superconducting order parameter.} \bl{It is useful to distinguish the relative and center-of-mass variables,
\begin{subequations}
\begin{align}
\label{eq.1.1}
& \drr 	= \frac{\dr + \dr'}{2},		& \bm{\rho}	&	= \dr - \dr',  \\
\label{eq.1.2}
& T 	= \frac{t + t'}{2}, 				& \tau		&	= t - t',
\end{align}
\end{subequations}
The relative coordinates \((\boldsymbol{\rho},\tau)\) characterize the internal structure of a Cooper pair, including its orbital and frequency parity, whereas the center-of-mass coordinates \((\mathbf R,T)\) characterize spatial or temporal modulation of the superconducting state itself.  }

\begin{figure}[t!]
\centering 
\includegraphics[width=\columnwidth]{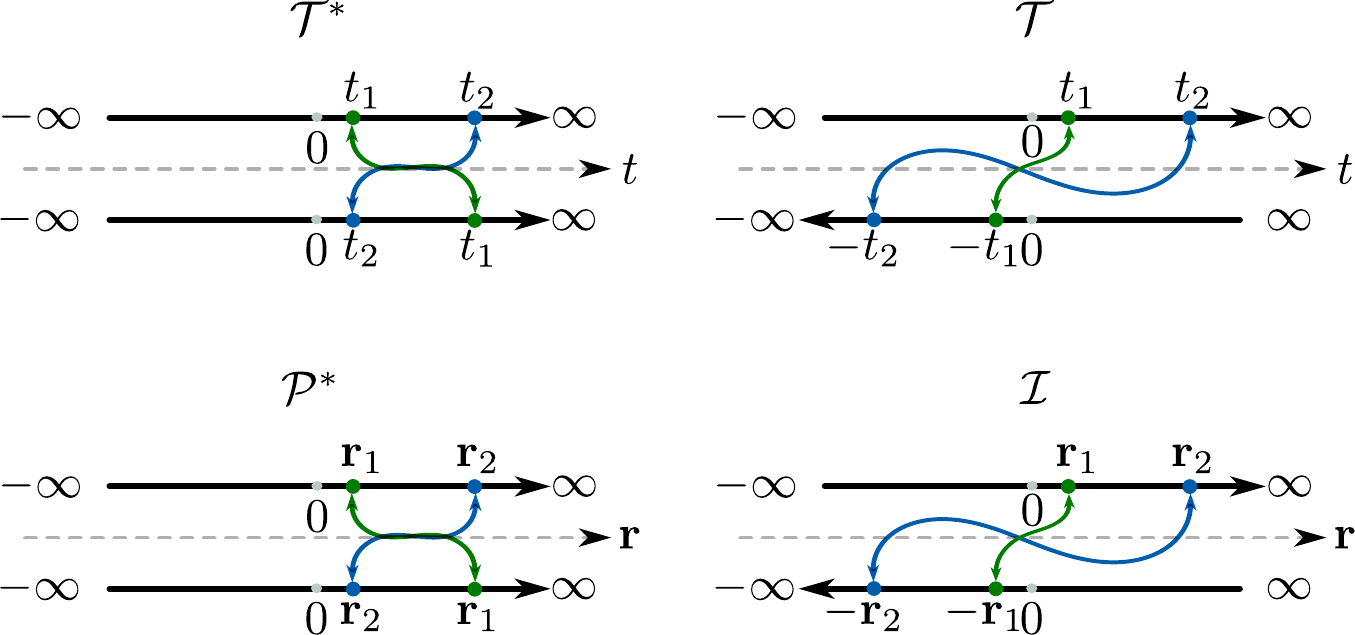}
\caption{Top: Schematic illustration of two symmetry operations: time \bl{swap} $\dt$ and physical time reversal $\cal T$. The operation $\dt$ interchanges, or swaps, the two time coordinates $t_1$ and $t_2$, whereas $\cal T$ maps $t \rightarrow -t$ and denotes the usual time-reversal operation. Bottom: Analogous distinction between the Cooper-pair spatial swap operation $\dpp$ and the physical inversion operation $\cal I$. \bl{The contour-ordered generalization of the time-sector operations, relevant for non-equilibrium settings with explicit center-of-mass time dependence, is shown in Fig.~\ref{fig:AFig1} and discussed in Appendix~\ref{sec.app.1}.}}
\label{fig:Fig1}
\end{figure}

The majority of the discussions on the nature of the superconducting state assume equal time pairing that can be perturbed away by allowing Eliashberg-like dependence of the superconducting order on time primarily through retardation effects. \bl{More general unequal-time pairing correlations have also attracted significant attention, most prominently in the context of odd-frequency (odd-\(\omega\)) superconductivity~\cite{RevModPhys.77.1321,PhysRevB.76.054522,Tanaka2012,RevModPhys.91.045005}. In contrast, the classification developed below treats dependence on \(\boldsymbol{\rho}\), \(\tau\), \(\mathbf R\), and \(T\) on equal footing, thereby organizing superconducting orders according to both the internal structure of the pair and the possible spatial or temporal modulation of the condensate. The odd-$\tau$ states can emerge in equilibrium. The states with nontrivial evolution in $T$, as defined above,  are  externally driven states and develop under Floquet drive, external voltage, light pulses etc. Hence nontrivial $T$ dependent states are driven dynamical states and we will discuss  new symmetry allowed states below. }

\bl{Here, we take a broader point of view and classify superconducting pairing states by the symmetry properties of their anomalous correlation functions. 
In this framework, conventional even-frequency (even-$\omega$) superconductors, odd-$\omega$ superconductors, finite-momentum Fulde--Ferrell--Larkin--Ovchinnikov (FFLO) or pair-density-wave (PDW) states, time-modulated superconductors, and their possible hybrid combinations are treated on equal footing. The role of symmetry is to identify which pairing structures are allowed by fermionic carriers; the actual realization of a given state is then determined by microscopic energetics, stability, disorder, and competing orders. Thus, odd-$\omega$ and time-modulated pairing are not exceptions to the usual classification. They are natural parts of the same broader symmetry framework. Similarly, more exotic states, such as odd-$\omega$ finite-momentum or time-modulated pairing, emerge as symmetry-allowed possibilities.}

We start by using the analysis proposed by Berezinskii~\cite{Berezinskii1974}. Fermi statistics of the electron operators, entering the pairing correlation function, introduce symmetry constraints on the swap properties of the two operators. Here, we introduce the relevant parity operators for the pair amplitude with respect to the relative spatial coordinates ($\dr$ to $\dr'$ swap) as $\dpp$: 
\begin{equation}\label{eq.2}
\dpp \df_{\alpha\beta,ab}(\dr, \dr'\,|\,t,t') {\dpp}^{-1} = \df_{\alpha\beta,ab}(\dr', \dr\,|\,t,t'),
\end{equation}
while the time-swap operator for the relative time is defined as $\dt$ \bl{, with its contour-ordered generalization discussed in Appendix~\ref{sec.app.1}}:
\begin{equation}\label{eq.3}
\dt \df_{\alpha\beta,ab}(\dr, \dr'\,|\,t,t') {\dt}^{-1} = \df_{\alpha\beta,ab}(\dr,\dr'\,|\, t',t).
\end{equation}
It is important to emphasize that the parity operators $\dpp$ and $\dt$ are distinct from the standard inversion operator $\mathcal{I}$ and the time-reversal operator $\mathcal{T}$. The former operate on relative coordinates, while the latter are defined using absolute spatial and temporal coordinates. In particular, applying time reversal to the pairing amplitude in Eq.~\eqref{eq.2} would require the Hermitian conjugate of the pairing field, which is not needed in the definitions associated with the time index swap $\dt$. The key difference between $\mathcal{T}$ and $\dt$ is illustrated in the top panel of Fig.~\ref{fig:Fig1}, which compares the time swap operator $\dt$ with the time reversal operator $\mathcal{T}$. A similar distinction is shown in the bottom panel of Fig.~\ref{fig:Fig1}, which depicts the Cooper pair spatial swap and inversion operators. \bl{For superconducting states with explicit center-of-mass time dependence, the same distinction can be formulated in terms of contour-ordered anomalous Green's functions on the Schwinger--Keldysh contour. In this formulation, $\dt$ swaps the two contour-time arguments of the anomalous correlator, whereas $\mathcal{T}$ acts as the physical time-reversal operation. The corresponding contour-based analysis is provided in Appendix~\ref{sec.app.1} (see Fig.~\ref{fig:AFig1} for visualization).} 

Finally, we define two other \bl{swap} operators, $\dss$, for the spin indices $\alpha,\beta$, and $\doo$, for the orbital indices $a,b$. For completeness, we show the action of these operators on the pairing correlation explicitly as
\begin{equation}\label{eq.4}
\dss \df_{\alpha\beta,ab}(\dr,\dr'\,|\,t,t') {\dss}^{-1} = \df_{\beta\alpha,ab}(\dr,\dr'\,|\,t,t'),
\end{equation}
while for the orbital permutation ($a$ to $b$ swap), one obtains 
\begin{equation}\label{eq.5}
\doo \df_{\alpha\beta,ab}(\dr,\dr'\,|\,t,t') {\doo}^{-1} = \df_{\alpha\beta,ba}(\dr,\dr'\,|\,t,t').
\end{equation}
Using only fermion commutation relations, it can be shown that the combined action of the individual \bl{swap} operators leads to a change in the sign of the pairing correlation function~\cite{RevModPhys.91.045005}, i.e., 
\begin{widetext}
\begin{equation}\label{eq.6}
\dss \dpp \doo \dt \df_{\alpha\beta,ab}(\dr,\dr'\,|\,t,t') {\dt}^{-1} {\doo}^{-1} {\dpp}^{-1} {\dss}^{-1} = - \df_{\alpha\beta,ab}(\dr,\dr'\,|\,t,t').
\end{equation}
\end{widetext}
or in a short notation: $\dss \dpp \doo \dt  = -1$. Since each of the permutation operators has two possibilities dictated by ${\dss}^2 = {\dpp}^2 = \doo^2 = {\dt}^2 = 1$ (assuming half-integer spin systems), there are $2^4$ total parities and half of them satisfy the required $\dss \dpp \doo \dt  = -1$ overall parity. Hence, we obtain a $2^3=8$, or {\em eight-fold}, \bl{swap}-symmetry classification for generic superconductors. We define odd-$\omega$ superconducting states as those with $\dt = -1$ symmetry, while even-$\omega$ states have $\dt = +1$ symmetry. This definition is consistent with the standard use of odd-$\omega$ pairing in terms of the anomalous Green's function~\cite{RevModPhys.91.045005}. 

\bl{For completeness we also include the classification in the presence of  spin-orbit coupling (SOC). In its absence, spin and orbital labels can be treated as independent quantum numbers. In this case, the spin-exchange parity $\dss$ distinguishes spin-singlet and spin-triplet channels, while $\doo$ distinguishes orbital-symmetric and orbital-antisymmetric channels. The constraint $\dss\dpp\doo\dt=-1$ then fixes the allowed combinations of spin, orbital, spatial, and relative-time parities. Thus, orbital degrees of freedom do not remove the singlet/triplet distinction; rather, they enlarge the set of symmetry-allowed pairing channels~\cite{Banerjee_2026}. In systems with strong SOC, however, spin and orbital angular momenta are generally not separately conserved. The classification should then be formulated in terms of the swap properties of the appropriate pseudospin or total angular momentum states, for example ${\bf J}={\bf L}+{\bf S}$, rather than in terms of separate spin and orbital parities. For example, in such cases one may use a generalized swap classification in which the separate spin and orbital parities are replaced by the swap of the relevant total angular momentum or pseudospin indices. Schematically, this amounts to replacing \(\dss\doo\) by an appropriate \({\cal J}\), leading to a constraint of the form \({\cal J}\dpp\dt=-1\)~\cite{PhysRevResearch.3.033255,PhysRevB.96.214514,PhysRevB.96.094526,Kim2018}. In principle, this can lead to more than eight-fold options, since $\cal J$ can be larger than 1/2.} 

It is straightforward to see that odd-$\omega$ states are distinct from their even-$\omega$ counterparts in both pure or hybrid cases: an odd-$\omega$ anomalous correlation function is odd under relative-time swap and therefore vanishes at equal relative time. \bl{This statement can be formulated equivalently in real time, imaginary time, or frequency space. The exchange-symmetry classification itself does not depend on the choice of representation. Real-time notation is useful for distinguishing the relative-time swap operation $\dt$ from the physical time-reversal operation $\mathcal T$, whereas Matsubara-frequency notation is often convenient for equilibrium superconductors because odd-$\omega$ pairing is then identified by the transformation $\omega_n\rightarrow -\omega_n$. For driven or genuinely non-equilibrium systems, the corresponding formulation should instead be written in terms of contour-ordered anomalous Green's functions on the Schwinger--Keldysh contour, as discussed in Appendix~\ref{sec.app.1}. For simplicity, hereafter we use the representation most convenient for the physical situation under discussion: equilibrium examples may be written in Matsubara frequency, while explicitly time-dependent or driven states are described in real-time or contour-time language.}

Considering a single-band (orbital) system with $\doo = +1$, a straightforward example in each case would be an even-$\omega$ ($\dt = +1$), $p$-wave (odd parity $\dpp = -1$), spin-triplet ($\dss = +1$) superconductor, as often discussed in connection with superfluid $^3$He states~\cite{PhysRevLett.29.1227} and was proposed to be the pairing state of a layered perovskite materials Sr$_2$RuO$_4$~\cite{Maeno1994}, although the $p$-wave pairing in Sr$_2$RuO$_4$ has been now challenged. In contrast, an odd-$\omega$ ($\dt = -1$), $s$-wave (even parity $\dpp = +1$), spin-triplet ($\dss = +1$) state has been theoretically proposed, for example, in two-band superconductors such as MgB$_2$ in the presence of an external magnetic field~\cite{PhysRevB.92.054516}. \bl{These examples illustrate how the same constraint $\dss\dpp\doo\dt=-1$ accommodates both conventional and Berezinskii-type pairing states without privileging a particular time representation.}

\bl{Finally, before turning to the geometric organization of the allowed pairing states following Eqs.~\eqref{eq.1.1}-\eqref{eq.1.2}, we clarify the relation between the anomalous Green's function ${\cal F}$ and the superconducting order parameter $\Delta$. In this work, the exchange-symmetry classification is formulated primarily in terms of ${\cal F}$, because this object is directly constrained by fermionic exchange statistics and is defined independently of a particular microscopic pairing mechanism. This choice is especially useful for relative-time-dependent, or odd-$\omega$ pairing states, for which the equal-time anomalous amplitude vanishes and the relevant information is contained in the finite-relative-time correlation function~\cite{RevModPhys.91.045005}. By contrast, a static order parameter of the type commonly used in Ginzburg--Landau descriptions is fully adequate for conventional static orders, but may obscure part of the relative-time structure of a dynamic pairing state. For this reason, ${\cal F}$ provides the most transparent object for the present symmetry classification.}

\subsection*{Anomalous Green's function or order parameter}

\bl{The anomalous Green's function and the superconducting order parameter are nevertheless closely related. In a general translationally invariant setting one may write, without committing to a particular microscopic pairing mechanism~\cite{PhysRevB.79.132502,PhysRevB.91.144514},
\begin{widetext}
\begin{equation}\label{eq.7}
\Delta_{\alpha\beta,ab}(i\omega_n,\vq) 
= 
-T \sum_{\omega_m,\vq'} 
{\cal D}_{\alpha\beta,ab}^{\gamma\delta,cd}
(i\omega_n,i\omega_m; \vq, \vq') 
{\cal F}_{\gamma\delta,cd}(i\omega_m,\vq'),
\end{equation}
\end{widetext}
where repeated spin and orbital indices are summed over. Here $(\alpha,\beta,\gamma,\delta)$ denote spin indices, $(a,b,c,d)$ denote orbital or band indices, $\Delta$ is the superconducting order parameter, and ${\cal D}_{\alpha\beta,ab}^{\gamma\delta,cd}
(i\omega_n,i\omega_m;\vq,\vq')$ is a general pairing kernel depending on Matsubara frequencies, momenta, and internal indices. We have written Eq.~\eqref{eq.7} in Matsubara notation for compactness, but the same relation can be formulated in real-frequency or contour-time language when appropriate.

It shows that the symmetry of $\Delta$ follows from the combined symmetry properties of ${\cal F}$ and the pairing kernel ${\cal D}$. Thus, once the transformation properties of ${\cal D}$ are specified, a classification formulated in terms of ${\cal F}$ can be translated into the corresponding classification of $\Delta$. In this sense, the two formulations are equivalent for the purpose of the symmetry classification. If one instead starts directly from $\Delta$, the symmetry properties of the pairing kernel must be specified. For example, in an effective-action treatment, the reality of the partition function imposes the hermiticity condition~\cite{PhysRevB.91.144514}
\begin{equation}\label{eq.8}
{\cal D}_{\alpha\beta,ab}^{\gamma\delta,cd}
(i\omega_n,i\omega_m; \vq, \vq')
=
\left[
{\cal D}^{\alpha\beta,ab}_{\gamma\delta,cd}
(i\omega_m,i\omega_n; \vq', \vq)
\right]^*,
\end{equation}
where $^{*}$ denotes complex conjugation. Such additional constraints make the order-parameter-based classification more involved, although it is ultimately equivalent once the pairing kernel is specified. Throughout the present work, we therefore use ${\cal F}$ as the primary object of the exchange-symmetry classification and reserve $\Delta$ for the superconducting order parameter. Related order-parameter-based formulations of odd-$\omega$ superconductivity have been discussed, for example, in Ref.~\cite{Kusunose2011}.}

After these preliminaries, we are in a position to illustrate the classification scheme based on the relative swap parities of space, time, spin, and orbital indices in the pairing channel. \bl{While the procedure is simple, it provides a unified organization of conventional, odd-frequency, finite-momentum, time-modulated, and hybrid superconducting orders.}

\section{Symmetries of the superconducting pairing \label{sec:sec.2}}

\begin{figure}[htb!]
\centering \includegraphics[width=0.8\linewidth]{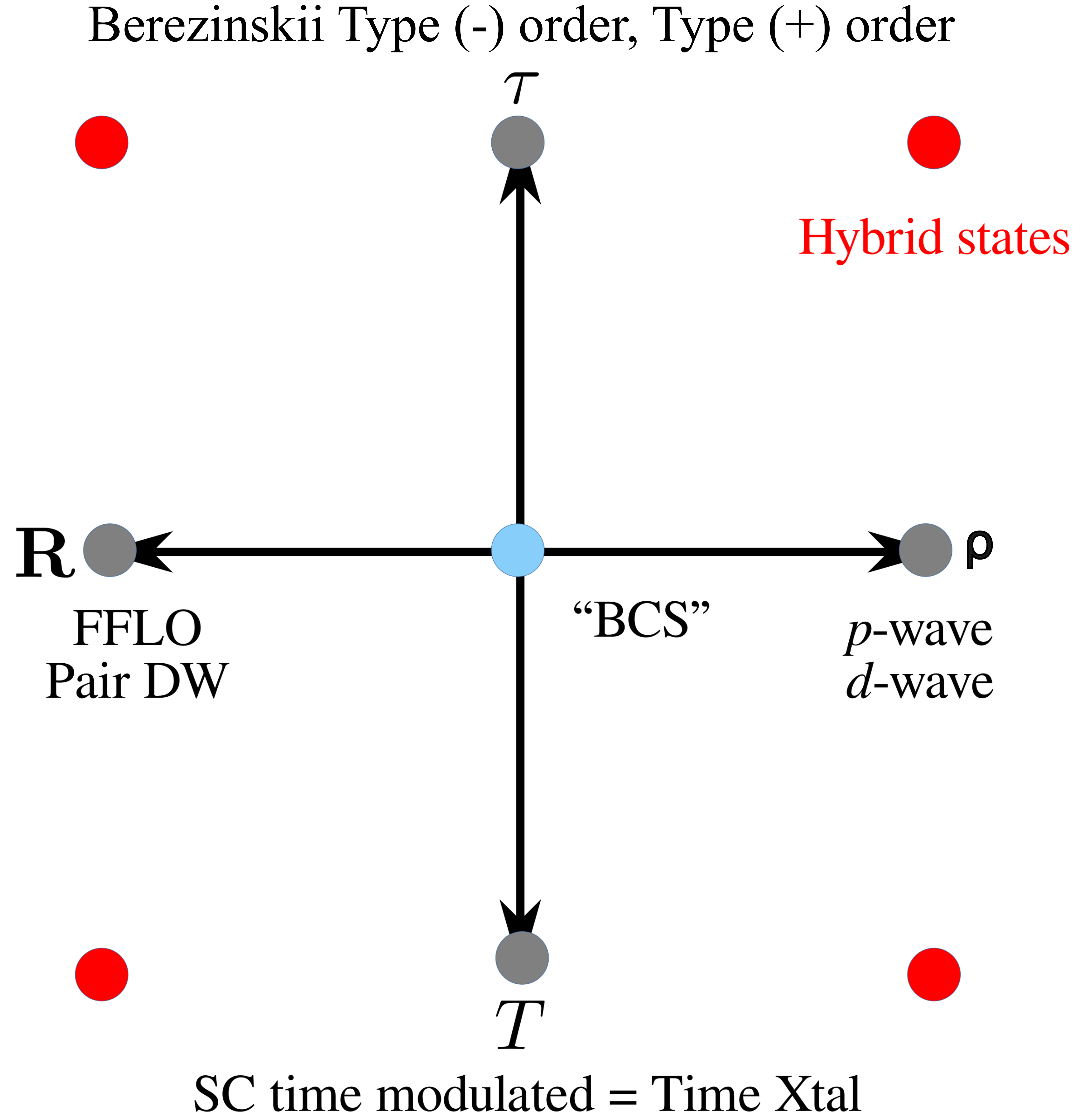}
\caption{\bl{Proposed Berezinskii--Abrahams (BA) hypercube for organizing superconducting anomalous correlation functions $\df_{\alpha\beta,ab}(\drr,\bm{\rho}|T,\tau)$ in the mixed relative/center-of-mass representation. The axes $\bm{\rho}$ and $\tau$ describe the internal spatial and relative-time structure of the Cooper pair, while $\drr$ and $T$ describe center-of-mass spatial and temporal modulation of the condensate. Since the four independent coordinates cannot be displayed as a true four-dimensional cube, the figure shows a schematic projection. Gray points indicate representative single-axis sectors: unconventional $p$- or $d$-wave pairing along $\bm{\rho}$, Berezinskii relative-time structure along $\tau$, FFLO/PDW order along $\drr$, and time-modulated superconductivity along $T$. Red points indicate representative hybrid sectors involving more than one nontrivial dependence, such as higher-angular-momentum odd-frequency pairing, odd-frequency FFLO/PDW order, unconventional time-modulated superconductivity, and time-modulated FFLO/PDW order. Here, Berezinskii Type $(-)$ denotes $\df(\tau)=-\df(-\tau)$, while Type $(+)$ denotes $\df(\tau)=\df(-\tau)$. The figure is qualitative; details of the axes and hybrid sectors are discussed in the main text.}}\label{fig:Fig2}
\end{figure}

\subsection{Berezinskii--Abrahams hypercube \label{sec:sec.2.1}}

\bl{As introduced earlier, the anomalous Green's function $\df_{\alpha\beta,ab}(\dr,\dr'\,|\,t,t')$ can be expressed in terms of the relative and center-of-mass variables $(\bm{\rho},\tau,\drr,T)$. The central idea of the present construction is to treat these four variables as independent axes for superconducting orders.} 
We illustrate the various possible superconducting orders in this four-dimensional phase space as shown in Fig.~\ref{fig:Fig2}. \bl{We refer to this organizing framework as the Berezinskii--Abrahams (BA) hypercube.}

The origin of the BA hypercube corresponds to the conventional uniform BCS $s$-wave superconductor, for which the anomalous correlation function has no explicit dependence on any of the four coordinates 
beyond the usual short-range structure of the Cooper pair. Moving along the relative spatial coordinate $\bm{\rho}$ introduces nontrivial internal orbital structure of the pair, corresponding to unconventional pairing channels such as $p$-, $d$-, or $f$-wave superconductivity. Moving along the relative-time coordinate $\tau$ corresponds to the even- or odd-$\omega$ classification originally introduced by Berezinskii~\cite{Berezinskii1974,PhysRevB.45.13125,RevModPhys.91.045005}. 

\bl{The remaining two axes describe center-of-mass structure rather than the internal structure of the pair.} An explicit center-of-mass spatial coordinate $\drr$-dependence leads to finite-momentum superconducting states, such as FFLO states~\cite{PhysRev.135.A550,Larkin1964} or PDW states~\cite{Daniel2020}. An explicit center-of-mass temporal coordinate $T$-dependence corresponds to time-modulated superconducting order, which may arise in driven, Floquet, voltage-biased, or time-crystalline settings~\cite{PhysRevLett.109.160401}. \bl{Thus, the four axes of the BA hypercube separate two types of structure: $({\bm{\rho}},\tau)$ characterize the internal space-time symmetry of the pair, while $(\drr,T)$ characterize spatial or temporal modulation of the condensate as a whole. Hybrid states correspond to sectors in which more than one of these structures is simultaneously present.}

\bl{It is important to emphasize that Fig.~\ref{fig:Fig2} organizes the space-time dependence of the anomalous correlation function; it does not by itself display the full spin and orbital exchange classification. The latter is encoded in the constraint $\dss\dpp\doo\dt=-1$ and summarized in Table~\ref{tab:Table_I}. Thus, a point or sector in the BA hypercube should be read together with the spin and orbital exchange parities. For example, an even-$\omega$, spin-singlet, $d$-wave FFLO state lies in the sector with nontrivial $\bm{\rho}$-dependence and nontrivial $\drr$-dependence, while remaining in the Berezinskii Type $(+)$ sector along the $\tau$-axis.}

\bl{To make this correspondence explicit, we discuss representative three-dimensional projections of the four-dimensional BA hypercube in Sec.~\ref{sec:sec.3}. Each projection displays a cubic sector containing various superconducting phases, candidate realizations, and symmetry-allowed hybrid states. This organization is intended to help readers connect familiar superconducting orders with the corresponding regions of the BA hypercube. }

The idea of a BA hypercube is motivated by the cube of theories controlled by the gravitational constant $G$, the speed of light $c$, and Planck's constant $\hbar$, often used in the high-energy and gravity communities to organize different limiting regimes of physical theories. \bl{In analogy with that construction, the BA hypercube is intended as an organizing diagram rather than a microscopic phase diagram. 
}

\subsection{Momentum-frequency representation \label{sec:sec.2.2}}

\bl{A useful way to make the four axes of the BA hypercube explicit is to Fourier transform both the relative and center-of-mass variables. Starting from the mixed representation $\df_{\alpha\beta,ab}(\drr,\bm{\rho}\,|\,T,\tau)$, we define}
\begin{widetext}
\begin{equation}\label{eq.9}
\bl{
\df_{\alpha\beta,ab}(\vQ,\vk\,|\,\Omega,\omega)
=
\int d\drr\, d\bm{\rho}\, dT\, d\tau\,
e^{-i\vQ\cdot\drr}
e^{-i\vk\cdot\bm{\rho}}
e^{i\Omega T}
e^{i\omega\tau}
\df_{\alpha\beta,ab}(\drr,\bm{\rho}\,|\,T,\tau).
}
\end{equation}
\end{widetext}
\bl{Here $\vk$ is the relative momentum conjugate to $\bm{\rho}$, $\omega$ is the relative frequency conjugate to $\tau$, $\vQ$ is the center-of-mass momentum conjugate to $\drr$, and $\Omega$ is the center-of-mass frequency conjugate to $T$. In this language, conventional uniform BCS pairing corresponds to the sector with $\vQ=0$, $\Omega=0$, and even $\omega$ structure. Unconventional $p$-, $d$-, or $f$-wave pairing is encoded in the dependence on $\vk$. Odd-$\omega$ pairing is encoded in the parity under $\omega \rightarrow -\omega$. FFLO or PDW order corresponds to finite $\vQ$, while time-modulated or driven superconducting order corresponds to finite $\Omega$. 
Hybrid states correspond to simultaneous nontrivial dependence on more than one of these variables.}

\bl{The Pauli principle acts on the relative variables, not on the center-of-mass variables. Therefore, for fixed $(\vQ,\Omega)$, exchange of the two fermions gives}
\begin{equation}\label{eq.10}
\bl{
\df_{\alpha\beta,ab}(\vQ,\vk\,|\,\Omega,\omega)
=
-
\df_{\beta\alpha,ba}(\vQ,-\vk\,|\,\Omega,-\omega).
}
\end{equation}
\bl{Eq.~\eqref{eq.10} is the compact momentum-frequency form of the exchange constraint. It shows that the eight-fold $\dss\dpp\doo\dt=-1$ classification concerns the parities under spin exchange, orbital exchange, $\vk \rightarrow -\vk$, and $\omega \rightarrow -\omega$. 
The uniform equilibrium limit used in many conventional discussions is recovered by setting $\vQ=0$ and $\Omega=0$. In this limit, Eq.~\eqref{eq.10} reduces to}
\begin{equation}\label{eq.11}
\bl{
\df_{\alpha\beta,ab}(\vk;\omega)
=
-
\df_{\beta\alpha,ba}(-\vk;-\omega).
}
\end{equation}
\bl{The familiar even- and odd-$\omega$ sectors then follow by assigning definite parity under $\omega \rightarrow -\omega$ ($\vk \rightarrow -\vk$), together with definite spin, and orbital parities.} 
\bl{In this regard, we note that conventional BCS theory or the Eliashberg framework for simple $s$-wave superconductors does not generically produce an odd-$\omega$ superconducting correlation function. This obstruction has been discussed as a ``no-go theorem" for odd-$\omega$ pairing~\cite{Langmann2022}. It does not rule out the generic odd-$\omega$ states, but  emphasizes the need for additional ingredients such as internal degrees of freedom, broken symmetries, interfaces, finite-momentum pairing, strong correlations, or non-equilibrium driving.}

\subsection{Swap classification and physical symmetries \label{sec:sec.2.3}}

Focusing on Eq.~\eqref{eq.6}, we can now tabulate all possible classes of superconductors based on their exchange parities in Table~\ref{tab:Table_I}. \bl{This table classifies the internal exchange structure of the pair in terms of spin, relative spatial parity, orbital parity, and relative-time parity. The BA hypercube in Fig.~\ref{fig:Fig2} complements this table by organizing the additional dependence on the center-of-mass space-time variables. Equivalently, Table~\ref{tab:Table_I} classifies the parities associated with $(\vk,\omega)$ and the internal spin/orbital labels, while the hypercube organizes how these states extend into sectors with finite $\vQ$, and $\Omega$, or both. Together, Table~\ref{tab:Table_I} and Fig.~\ref{fig:Fig2} provide the symmetry roadmap used in the rest of the paper.}

\begin{table}[t!]
\centering
\begin{tabular}{|c|c|c|c|c|c|c|c|c|}
\hline
Operator & \multicolumn{4}{|c|}{\textbf{Static} order Type (+) } & \multicolumn{4}{|c|}{\textbf{Dynamic} order Type (-)} \\ \hline
$\dss$	& \mb{$-$}	& \mb{$-$}	& \mb{$+$}	& \mb{$+$}	& \mb{$+$}	& \mb{$+$}	& \mb{$-$}	& \mb{$-$}	\\ \hline
$\dpp$	& \mb{$+$}	& \mb{$-$}	& \mb{$+$}	& \mb{$-$}	& \mb{$+$}	& \mb{$-$}	& \mb{$+$}	& \mb{$-$}	\\ \hline
$\doo$	& \mb{$+$}	& \mb{$-$}	& \mb{$-$}	& \mb{$+$}	& \mb{$+$}	& \mb{$-$}	& \mb{$-$}	& \mb{$+$}	\\ \hline
$\dt$	& \mb{$+$}	& \mb{$+$}	& \mb{$+$}	& \mb{$+$}	& \mb{$-$}	& \mb{$-$}	& \mb{$-$}	& \mb{$-$}	\\	\hline
\end{tabular}
\caption{Eight-fold classification of superconductors based on the permutation symmetries for the two fermion operators in the Cooper pair correlation function. Note that the signature, namely the product of all parities in each column, remains $-1$, reflecting $\dss \dpp\doo\dt = -1$. For convenience, we use the notation $(\dss \dpp \doo \dt)$ to label the states in each column. The BCS state, shown in column 1, corresponds to $(-+++)$. Other states are labeled similarly~\cite{RevModPhys.91.045005}.}
\label{tab:Table_I}
\end{table}

In addition to the $\dss \dpp \doo \dt$ classification, time-reversal ($\mathcal{T}$) and spatial inversion ($\mathcal{I}$) operations open up the more possibilities. Note that $\mathcal{T}$ ($\mathcal{I}$) changes the sign of the center-of-mass coordinate and the relative coordinates as follows
\begin{equation}\label{eq.12}
\mathcal{T} : \{ T, \tau\} \rightarrow \{-T, -\tau\}; \quad
\mathcal{I} : \{\drr, \bm{\rho} \} \rightarrow \{-\drr, -\bm{\rho}\}.
\end{equation}
\bl{It is important to contrast Eq.~\eqref{eq.12} with swap operations $\dt$ and $\dpp$ introduced earlier:
\begin{equation}
\dt: \{T, \tau\} \rightarrow \{ T, -\tau\}; \quad
\dpp: \{\drr,\dr\} \rightarrow \{\drr, -\dr\}
\end{equation}
In momentum space, $\cal I$ reverses momentum but not spin, while $\cal T$ reverses both momentum, spin, and orbital angular momenta ${\cal T}: \{\sigma,l\}\rightarrow\{-\sigma,-l\}$. On the other hand, $\mathcal{I}$ leaves these angular momenta unchanged, ${\cal I}: \{\sigma,l\}\rightarrow\{\sigma,l\}$. This distinction between exchange operations and global symmetry operations is central to the classification. Taken together, the BA hypercube and the $\dss \dpp \doo \dt$ classification provide a symmetry framework for organizing pure and hybrid superconducting orders. In the following section, we discuss representative examples, including possible microscopic origins, candidate platforms, and experimental signatures. The stability issues are discussed in Sec. V.}


\begin{figure*}[t!]
\centering
\includegraphics[width=0.95\linewidth]{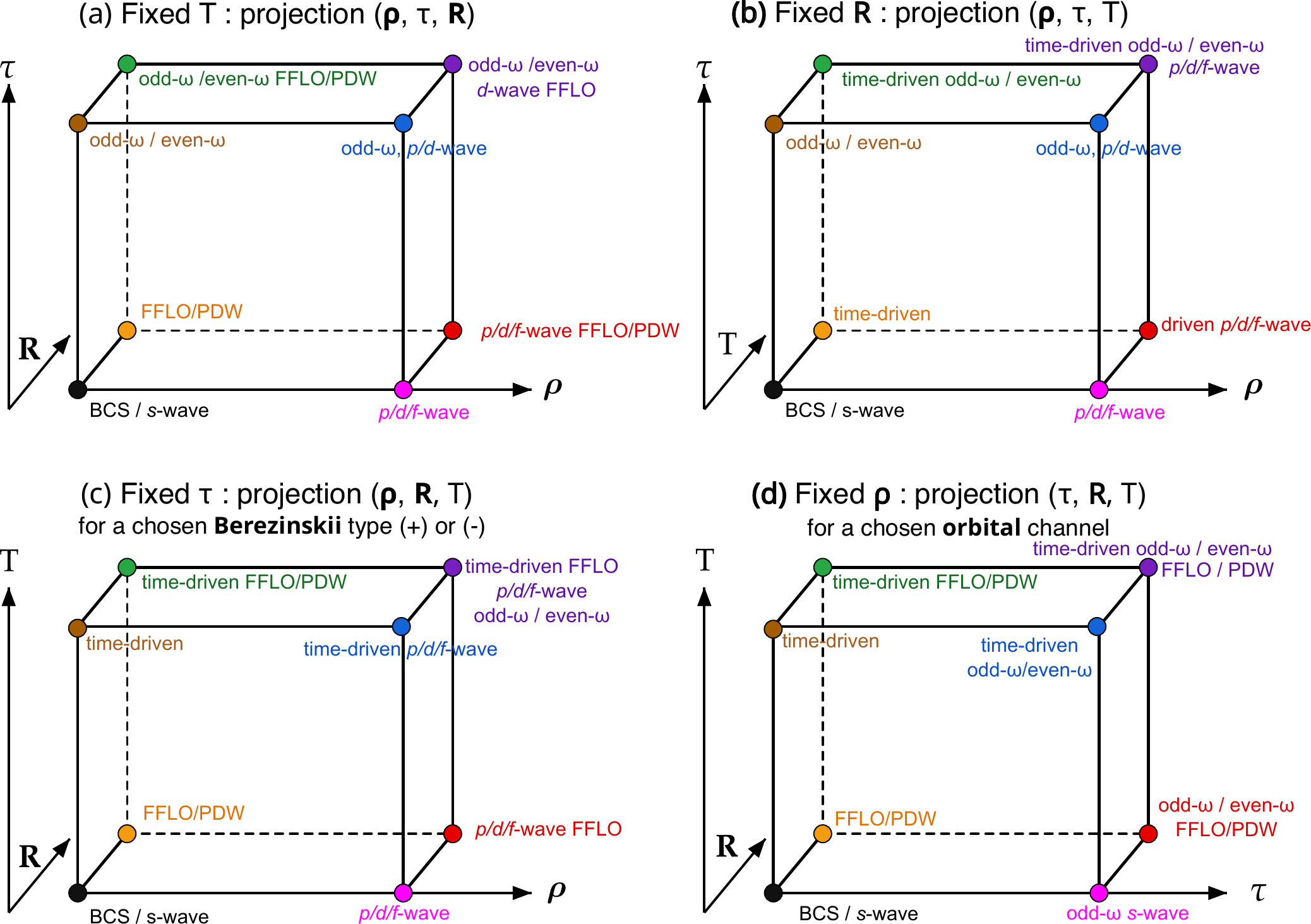}
\caption{\bl{Four representative three-dimensional projections of the Berezinskii--Abrahams hypercube obtained by fixing one of the four coordinates $(\bm{\rho},\tau,\drr,T)$: (a) fixed $T$, giving the $(\bm{\rho},\tau,\drr)$ cube; (b) fixed $\drr$, giving the $(\bm{\rho},\tau,T)$ cube; (c) fixed $\tau$, giving the $(\bm{\rho},\drr,T)$ cube; and (d) fixed $\bm{\rho}$, giving the $(\tau,\drr,T)$ cube. In each case, representative known superconducting states and symmetry-allowed hybrid sectors are indicated schematically. The detailed meaning of each projection is discussed in Sec.~\ref{sec:sec.3}.}}\label{fig:FigRoadmap}
\end{figure*}

\section{Roadmap of BA-hypercube sectors \label{sec:sec.3}}

\bl{
Here, we analyze representative three-dimensional projections of the BA hypercube. Choosing any three coordinates out of the four variables $(\bm{\rho},\tau,\drr,T)$ produces a cubic projection of the full BA hypercube. There are therefore four natural three-dimensional projections, corresponding to fixing one of the four coordinates in turn. These projections 
represent a symmetry-based  connections between known superconducting orders. For clarity, the corresponding schematic cubes are displayed in Fig.~\ref{fig:FigRoadmap}, while more detailed projection diagrams   are discussed in the following sections. In this way, cubes make the abstract four-dimensional organization of Fig.~\ref{fig:Fig2} more transparent and provide a direct bridge between the classification scheme and the microscopic examples discussed in the following sections.}

\subsection{Projection I : \texorpdfstring{$(\bm{\rho},\tau,\drr)$}{(rho,tau,R)} at fixed \texorpdfstring{$T$}{T}}

\bl{Fixing the center-of-mass time $T$ gives the $(\bm{\rho},\tau,\drr)$ projection shown in Fig.~\ref{fig:FigRoadmap}(a). This is the natural equilibrium 
cube: $\bm{\rho}$ describes internal orbital structure, $\tau$ describes the Berezinskii Type $(+)$ or $(-)$ relative-time sector, and $\drr$ describes finite center-of-mass momentum order, equivalently finite $\vQ$. Thus the origin corresponds to a uniform $s$-wave BCS superconductor~\cite{PhysRev.108.1175}, the $\bm{\rho}$-axis to unconventional $p$-, $d$-, or $f$-wave pairing~\cite{Maeno1994}, the $\tau$-axis to odd-$\omega$ superconductivity~\cite{Berezinskii1974,PhysRevB.45.13125,RevModPhys.91.045005}, and the \(\drr\)-axis to FFLO or PDW order~\cite{PhysRev.135.A550,Larkin1964,Daniel2020}.}

\bl{The mixed vertices collect hybrid equilibrium sectors: higher-angular-momentum odd-$\omega$ pairing in the $(\bm{\rho},\tau)$ plane~\cite{PhysRevB.104.094506}, finite-angular momentum FFLO/PDW states in the $(\bm{\rho},\drr)$ plane~\cite{PhysRevLett.96.117006}, and odd-$\omega$ FFLO/PDW correlations in the $(\tau,\drr)$ plane~\cite{Chakraborty_2021,PhysRevLett.129.177601}. An exotic example in this case would be : non-$s$-wave, odd-$\omega$, FFLO/PDW state. A closely related example with $p$-wave sub-dominant odd-$\omega$ PDW state was studied by Yoshida et al.~\cite{PhysRevB.104.094506}, but pure odd-$\omega$ with the additional structures [cf. Fig.~\ref{fig:FigRoadmap}(a)] is not analyzed to the best of our knowledge. Additionally, the stability of various sectors is a separate microscopic question, especially when odd-$\omega$ structure is combined with finite-$\vQ$ order.}

\subsection{Projection II : \texorpdfstring{$(\bm{\rho},\tau,T)$}{(rho,tau,T)} at fixed \texorpdfstring{$\drr$}{R}}

\bl{Fixing the center-of-mass coordinate $R$, or equivalently setting $Q=0$, gives the $(\rho,\tau,T)$ projection shown in Fig.~3(b). This is the spatially uniform but time-dependent cube: $\rho$ labels the internal orbital structure, $\tau$ labels the Berezinskii Type $(+)$ or Type $(-)$ relative-time sector, and $T$ labels explicit center-of-mass time dependence, equivalently finite center-of-mass frequency $\Omega$. The origin is the uniform $s$-wave BCS state~\cite{PhysRev.108.1175}; the $\rho$-axis gives unconventional $p$-, $d$-, or $f$-wave pairing~\cite{Maeno1994}; the $\tau$-axis gives odd-$\omega$ superconductivity~\cite{Berezinskii1974,PhysRevB.45.13125,RevModPhys.91.045005}; and the $T$-axis gives time-modulated superconducting order, as may arise in voltage-biased Josephson junctions, driven/Floquet superconductors, pump-probe settings, or time-crystalline superconducting states~\cite{Josephson1962,PhysRevB.80.205408,PhysRevB.103.104505,Dehghani2021,Kitamura2022,PhysRevLett.109.160401,Sacha2018,Else2020}.

This projection emphasizes that odd-$\omega$ structure and time modulation are distinct. Odd-$\omega$ pairing is odd under relative-time exchange ($\omega\rightarrow-\omega$), whereas nontrivial $T$-dependence describes a condensate whose center-of-mass phase or amplitude evolves in time. In a driven or Floquet setting this may be represented schematically as
\begin{equation}\label{eq.Floquet_Orbital}
c(T,\tau)=\sum_N e^{-iN\Omega_{\rm D}T}c_N(\tau),    
\end{equation}
where $c$ is the electron annihilation operator, $\Omega_{\rm D}$ is the drive frequency; and the $N$ plays the role of an orbital index. The parity of each ${\cal F}_{NN'}(\tau)$ under $\tau\rightarrow-\tau$ then determines whether that component is even- or odd-$\omega$. The mixed vertices describe spatially homogeneous hybrid dynamical sectors: higher-angular-momentum odd-$\omega$ pairing in the $(\rho,\tau)$ plane~\cite{PhysRevB.104.094506}, unconventional time-modulated superconductivity in the $(\rho,T)$ plane, and driven even- or odd-$\omega$ superconductivity in the $(\tau,T)$ plane~\cite{PhysRevB.103.104505,PhysRevB.90.205127}. The fully mixed corner corresponds to a non-$s$-wave, odd-$\omega$, time-modulated superconducting state. This sector combines ingredients known separately from non-$s$-wave odd-$\omega$ pairing and Floquet-induced odd-$\omega$ superconductivity~\cite{RevModPhys.91.045005,PhysRevB.104.094503,PhysRevB.103.104505}, but its realization as a stable phase remains unexplored. In particular, genuinely time-modulated superconducting order should be understood in driven or non-equilibrium systems rather than as a generic thermal-equilibrium phase~\cite{Watanabe2015,Sacha2018,Else2020}.}

\subsection{Projection III : \texorpdfstring{$(\bm{\rho},\drr,T)$}{(rho,R,T)} at fixed \texorpdfstring{$\tau$}{tau}}

\bl{Fixing the relative-time sector gives the $(\bm{\rho},\drr,T)$ projection shown in Fig.~\ref{fig:FigRoadmap}(c). Since the $\tau$ direction encodes the Berezinskii Type $(+)$ or Type $(-)$ character, this projection should be viewed as two related cubes: one for even relative-time parity and one for odd relative-time parity. Within either cube, $\bm{\rho}$ labels the internal orbital structure of the Cooper pair, $\drr$ labels finite center-of-mass momentum or FFLO/PDW order, and $T$ labels explicit center-of-mass time dependence, equivalently finite center-of-mass frequency $\Omega$.}

\bl{In the Type $(+)$ cube, the main sectors are BCS superconductivity, higher angular-momentum pairing, FFLO/PDW order~\cite{PhysRev.135.A550,Larkin1964,PhysRevLett.96.117006,Daniel2020}, and time-modulated even-$\omega$ superconductivity~\cite{Josephson1962,PhysRevLett.109.160401,Watanabe2015,Sacha2018,Else2020}. The Type $(-)$ cube contains the corresponding odd-$\omega$ sectors, including odd-$\omega$ FFLO/PDW correlations~\cite{Chakraborty_2021,PhysRevLett.129.177601} and Floquet-engineered odd-$\omega$ superconductivity~\cite{PhysRevB.103.104505,Chakraborty_2022}. Thus this projection separates the choice of relative-time parity from the independent question of whether the condensate also carries orbital structure, finite momentum, time modulation, or combinations thereof.}

\bl{The mixed vertices represent hybrid sectors within a fixed Berezinskii class. For example, the $(\bm{\rho},\drr)$ plane contains unconventional FFLO/PDW states, the $(\bm{\rho},T)$ plane contains driven $p$-, $d$-, or higher-angular-momentum superconducting states, and the $(\drr,T)$ plane contains time-modulated FFLO/PDW order. The fully mixed corner corresponds to a finite-momentum, time-modulated, non-$s$-wave superconducting state, considered either in the Type $(+)$ or Type $(-)$ relative-time sector.} 

\subsection{Projection IV : \texorpdfstring{$(\tau,\drr,T)$}{(tau,R,T)} at fixed \texorpdfstring{$\bm{\rho}$}{rho}}

\bl{Fixing the relative spatial sector gives the $(\tau,\drr,T)$ projection shown in Fig.~\ref{fig:FigRoadmap}(d). Physically, this corresponds to choosing an orbital channel, for example an $s$-wave or a higher-angular-momentum channel, and then asking how relative-time parity, finite center-of-mass momentum, and center-of-mass time dependence combine within that channel.}

\bl{For an $s$-wave channel, the origin is the conventional BCS state, the $\tau$-axis gives odd-$\omega$ $s$-wave superconductivity, the $\drr$-axis gives FFLO/PDW order, and the $T$-axis gives time-modulated superconductivity. Odd-$\omega$ $s$-wave correlations are among the most robust odd-frequency components and are widely discussed in proximity structures and multiband systems~\cite{RevModPhys.77.1321,RevModPhys.91.045005,PhysRevB.88.104514,Triola2020}. The mixed sectors include odd-$\omega$ FFLO/PDW correlations~\cite{Chakraborty_2021}, driven odd-$\omega$ superconductivity~\cite{PhysRevB.103.104505}, and a fully mixed $(\tau,\drr,T)$ sector that remains unexplored.

The cubic projections in Fig.~\ref{fig:FigRoadmap} repeatedly point to two particularly unexplored sectors of the BA hypercube: an odd-$\omega$ PDW state and its driven analogue. We note that odd-$\omega$ PDW correlations have been discussed previously in underdoped cuprates in proximity to charge-density-wave order~\cite{Chakraborty_2021}, and that related finite-momentum odd-$\omega$ pairing appears in the context of Majorana Fermi surfaces~\cite{PhysRevLett.129.177601}. However, both examples remain difficult to isolate experimentally. In the cuprate setting, the odd-$\omega$ PDW component coexists with an even-$\omega$ counterpart, making an unambiguous identification challenging. In the Majorana-based proposal, the construction is theoretically compelling but relies on assumptions whose experimental realization has yet to be firmly tested. In Sec.~\ref{sec.odd-f-PDW}, we therefore propose a more direct and experimentally motivated route: a spin-active superconductor--half-metal interface, inspired by the Eschrig-L\"{o}fwander~\cite{Eschrig2008} mechanism, supplemented by spatially modulated and, in the driven case, time-dependent interface hopping. This setup is designed to generate a finite-$\bf Q$ odd-$\omega$ anomalous component in the half-metal side, whose structure can be analyzed explicitly by solving the corresponding Usadel equations~\cite{Bennemann2008}.}


\begin{table*}[htb!]
\caption{\bl{Representative microscopic routes, candidate platforms, and experimental probes for  Berezinskii--Abrahams hypercube. The listed mechanisms illustrate how additional microscopic structure, broken symmetry, or external driving can move a conventional superconducting state into another symmetry-allowed sector. SC denotes superconductivity, FM ferromagnet, STM scanning tunneling microscopy, NMR nuclear magnetic resonance, and ARPES angle-resolved photoemission spectroscopy.}}\label{tab:micro_routes}
\begin{ruledtabular}
\begin{tabular}{p{2.3cm} p{3.0cm} p{3.7cm} p{3.1cm} p{3.0cm}}
\textbf{route} & \textbf{main ingredient} & \textbf{representative sector} & \textbf{candidate platforms} & \textbf{possible probes} \\ \\
\hline

interface/proximity &
broken spin rotation; magnetic texture; spin-active scattering &
odd-$\omega$ spin-triplet correlations from even-$\omega$ spin-singlet superconductivity~\cite{RevModPhys.77.1321,PhysRevB.75.134510,PhysRevB.76.054522,Eschrig2008} &
SC/FM junctions; half-metals; spin-active interfaces &
Josephson current; spin susceptibility; tunneling spectroscopy \\ \\

Josephson/multi-terminal &
left-right lead index; phase bias; broken L-R symmetry &
odd-$\omega$ interlead anomalous correlations, e.g. $(-\!+\!++) \rightarrow (-\!+\!--)$~\cite{Balatsky2018,PhysRevB.92.024512} &
Josephson junctions; rings; multi-terminal devices &
phase-sensitive Josephson response; ac Josephson spectroscopy; flux interference \\ \\

finite-$\vQ$ route &
spatial modulation; band folding &
FFLO/PDW order~\cite{PhysRev.135.A550,Larkin1964,Daniel2020}; odd-$\omega$ FFLO/PDW hybrid sectors &
Pauli-limited SCs; cuprates; PDW systems; heterostructures &
STM; NMR; gap modulation; finite-$\vQ$ spectroscopy \\ \\

driven/Floquet route &
finite $\Omega$; Floquet sidebands; explicit $T$-dependence &
time-modulated superconductivity; driven odd-$\omega$ pairing~\cite{Josephson1962,PhysRevB.80.205408,PhysRevLett.109.160401,Watanabe2015,Sacha2018,Else2020,PhysRevB.103.104505,Dehghani2021,Kitamura2022} &
Floquet SCs; voltage-biased junctions; pump-probe platforms &
pump-probe spectroscopy; time-resolved ARPES; sidebands; ac response \\ \\

multiband / topological route &
band, orbital, chirality, pseudospin, or SOC degrees of freedom &
odd-$\omega$ multiband pairing; inter-chirality Dirac/Weyl pairing; SOC-enabled higher-$J$ sectors~\cite{PhysRevB.88.104514,PhysRevB.92.094517,Triola2020,PhysRevB.92.054516,PhysRevB.100.180502,Parhizgar2020,PhysRevB.100.104511,PhysRevResearch.3.033255,PhysRevB.96.214514,PhysRevB.96.094526,Kim2018} &
MgB$_2$; Dirac/Weyl semimetals; spin-orbit-coupled SCs; spin-$3/2$ systems &
interband response; surface spectroscopy; Josephson signatures; quasiparticle interference \\ \\

non-Hermitian / dissipative route &
spin-dependent decay; gain/loss imbalance; exceptional spectra &
odd-$\omega$ components induced or enhanced in open superconducting systems~\cite{Bandyopadhyay2020,PhysRevB.105.094502} &
dissipative SCs; open heterostructures; non-Hermitian platforms &
spectral asymmetry; lifetime-resolved spectroscopy; exceptional-point signatures \\ \\

Majorana/strong-correlation route &
self-conjugate fermions; fractionalization; strong correlations &
intrinsic odd-$\omega$ correlations in Majorana, SYK-like, or Kondo-based settings~\cite{PhysRevB.92.121404,Witten_2019,PhysRevD.94.106002,PhysRevB.49.8955} &
Majorana wires; SYK-like systems; Kondo lattices &
zero-bias tunneling; local correlation spectroscopy; anomalous dynamical response \\ 
\end{tabular}
\end{ruledtabular}
\end{table*}

\section{Microscopic routes, candidate platforms, and probes\label{sec:micro}}

\bl{The purpose of this section is to connect the geometric classification to microscopic mechanisms, candidate platforms, and experimental probes. We therefore organize representative realization routes according to the additional physical ingredients that move a superconducting system away from the conventional uniform BCS solutions. These ingredients include broken spin rotation, orbital or band structure, left-right lead indices, finite center-of-mass momentum, Floquet sidebands, dissipation, non-Hermitian dynamics, topology, and strong correlations. A compact summary of these routes, together with representative sectors, candidate platforms, and possible probes, is provided in Table~\ref{tab:micro_routes}.}

\subsection{General design principle}

\bl{The $\dss \dpp \doo \dt$ classification in Table~\ref{tab:Table_I} provides a practical rule for constructing candidate pairing sectors. An even-$\omega$, spin-singlet, even-parity superconducting state belongs to the $(-+++)$ sector.  The general $\dss \dpp \doo \dt$ classification on swap and permutation of relative indices remains valid for the pairing sates explicitly dependent on modulation of in $T$ an in ${\bf R}$.  This label system  generated by $\dss \dpp \doo \dt$ classification is agnostic of any center or mass spatial/temporal coordinates changes.}


\bl{The process to infer the potential for generation of new even-$\omega$  and odd-$\omega$ states is as follows:}
\begin{enumerate}
\item \bl{identify the parent pairing channel, usually an even-$\omega$, spin-singlet, even-parity BCS-like state.}
\item \bl{Identify the additional microscopic label or broken symmetry, such as magnetic order or spin-active interfaces~\cite{RevModPhys.77.1321,Eschrig2008}, spin-orbit coupling~\cite{PhysRevResearch.3.033255,Kim2018}, orbital or band structure~\cite{PhysRevB.88.104514,Triola2020}, lead asymmetry~\cite{Balatsky2018}, finite momentum or spatial modulation~\cite{PhysRev.135.A550,Larkin1964,Daniel2020,Chakraborty_2021,Chakraborty_2022}, Floquet sidebands~\cite{PhysRevB.103.104505}, dissipation or non-Hermitian structure~\cite{PhysRevB.105.094502}, or topological/Majorana degrees of freedom~\cite{RevModPhys.83.1057,PhysRevB.92.121404,PhysRevB.100.180502}.}
\item \bl{determine which exchange parity is changed or mixed by this ingredient, and hence which $\dss \dpp \doo \dt = -1$ constraint-allowed sector becomes accessible.}
\item \bl{determine whether the resulting object is a stable bulk phase, an induced anomalous correlation, or a transient/driven response.}
\end{enumerate}

\bl{The last step is essential. The BA hypercube specifies what is allowed by exchange symmetry, not what is guaranteed to occur. Many odd-$\omega$ or hybrid components appear naturally as induced anomalous correlations near interfaces, in multiband systems, or under drive. Establishing them as stable bulk phases requires additional energetic, disorder, and phase-stiffness criteria~\cite{Heid1995,c57s-skv9}. In this discussion we distinguish below between symmetry allowance, induced correlations, and stable realizable  odd-$\omega$ superconducting order.} \\

\subsection{Exchange-sector routes: spin, orbital, lead, and internal labels \label{sec.ex_sec}}


\bl{We first consider mechanisms that change the exchange structure of the anomalous Green's function. Magnetic textures or spin-active interfaces can mix singlet and triplet components, while orbital, band, chirality, or lead indices can provide an additional exchange degree of freedom. Once these indices are specified, the constraint $\dss\dpp\doo\dt=-1$ fixes the allowed even- and odd-$\omega$ pairing components.
}
\subsubsection{Interface dominated design}

\bl{A canonical example is provided by superconductor--ferromagnet and superconductor--half-metal heterostructures. When a conventional even-$\omega$, spin-singlet superconductor is coupled to a magnetic material or a spin-active interface, spin-rotation symmetry is broken and singlet correlations can be converted into triplet correlations. In diffusive proximity systems, the long-range triplet response is commonly associated with odd-$\omega$ $s$-wave triplet components, while even-$\omega$ $p$-wave triplet components and higher partial waves become important toward the ballistic regime~\cite{RevModPhys.77.1321,PhysRevB.75.134510,PhysRevB.76.054522,Eschrig2008}. Thus, the magnetic interface should be viewed as a mechanism for mixing spin-exchange sectors, not simply as a source of odd-$\omega$ pairing.}

\begin{figure}[t!]
\centering \includegraphics[width=1.0\linewidth]{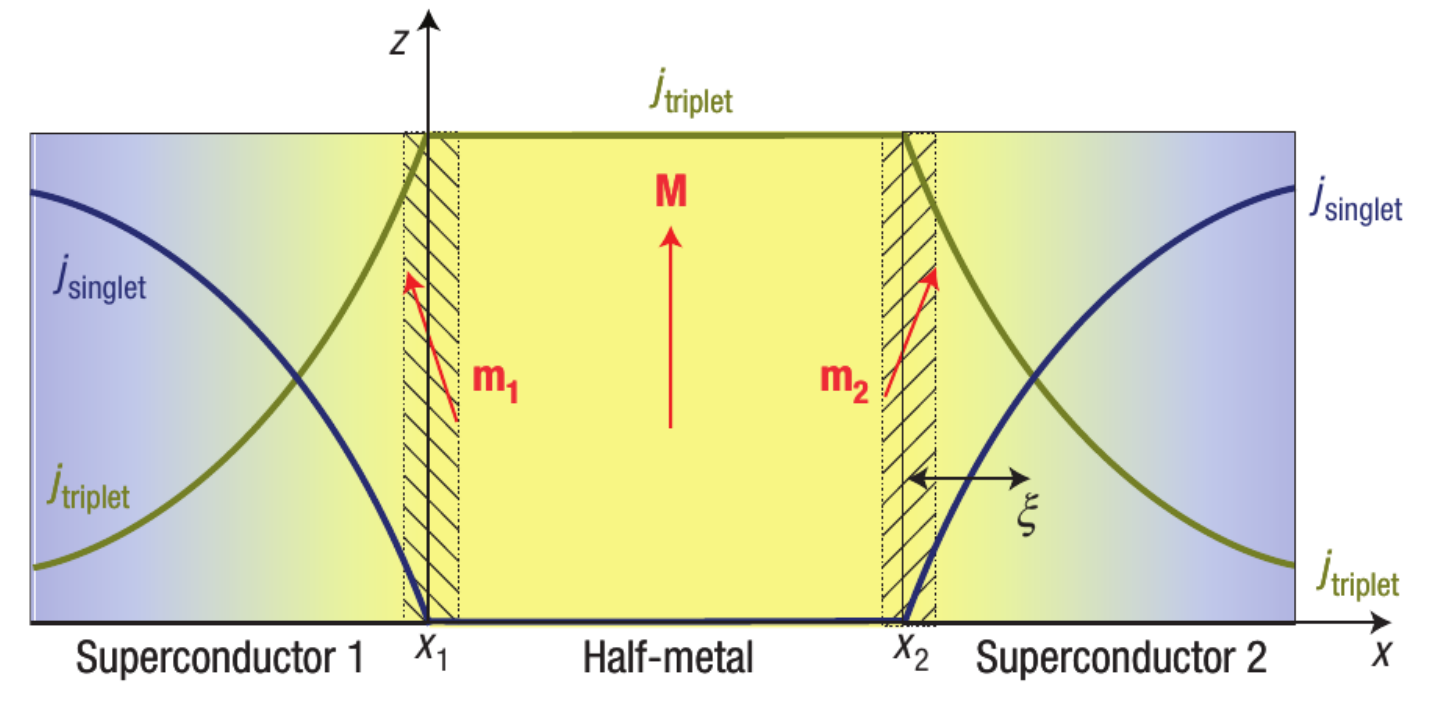}
\caption{\bl{Spin-active conversion of singlet supercurrent into equal-spin triplet supercurrent in a superconductor--half-metal junction. Misaligned interface moments $\mathbf{m}_1$ and $\mathbf{m}_2$ break spin-rotation symmetry with respect to the half-metal magnetization $\mathbf{M}$, allowing singlet correlations to be converted near the interfaces into long-range equal-spin triplet correlations. Only the triplet component propagates through the half-metal region. Adapted from Eschrig and L\"{o}fwander~\cite{Eschrig2008}.}}\label{fig:FigProximity}
\end{figure}

\bl{A microscopic illustration is the spin-active-interface theory of Eschrig and L\"{o}fwander~\cite{Eschrig2008}, shown schematically in Fig.~\ref{fig:FigProximity}. At a singlet-superconductor/half-metal interface, spin-dependent scattering first generates an $S=1$, $m=0$ triplet component. If spin-rotation symmetry is broken at the interface, spin-flip transmission converts this component into an equal-spin $S=1$, $m=1$ triplet amplitude, which can propagate through the half metal. In the tunneling limit, this equal-spin component has the schematic form
\begin{subequations}
\begin{align}
\label{eq:SF_1}
\df_{\uparrow\uparrow,j}^{s,a}(x) & \propto A_j |\Delta| e^{i\bar\chi_j} \frac{|\epsilon_n|}{\epsilon_n^2+|\Delta|^2} \Psi^{s,a}_{j}(x), \\
A_j & = 2t_{\uparrow\uparrow,j}t_{\downarrow\uparrow,j} \sin\frac{\vartheta_j}{2}. 
\end{align}
\end{subequations}
Here $\vartheta_j$ is the spin-mixing angle, while $t_{\uparrow\uparrow,j}$ and $t_{\downarrow\uparrow,j}$ are spin-conserving and spin-flip transmission amplitudes at interface $j$. The amplitude therefore requires both spin mixing and spin-rotation-symmetry breaking.} 

\begin{figure}[t!]
\centering
\includegraphics[width=0.7\linewidth]{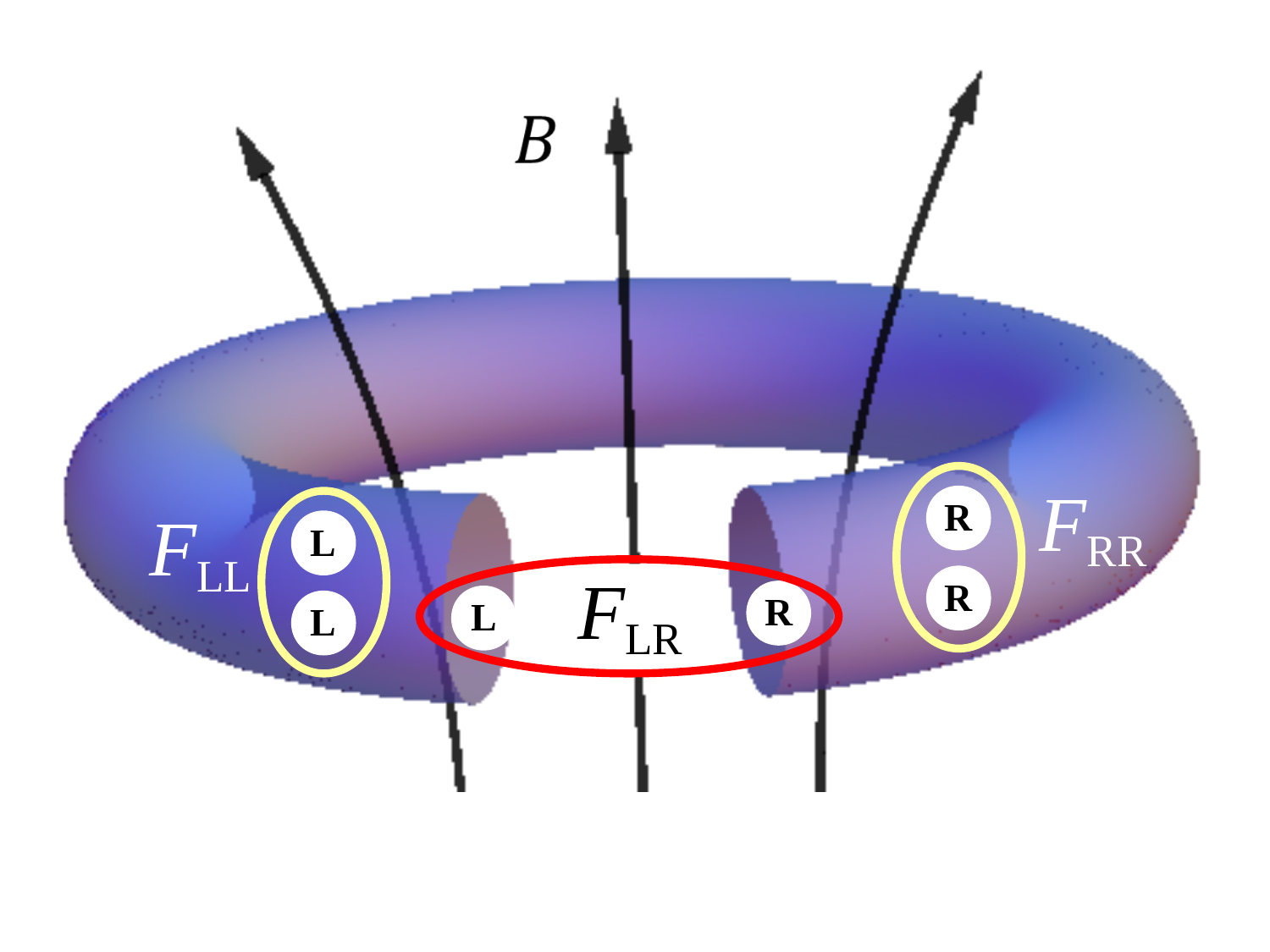}
\caption{\bl{Induction of interlead superconducting correlations in a Josephson junction in a ring geometry. If the phase difference $\phi_{\rm LR}=\phi_{\rm L}-\phi_{\rm R}$, controlled by an external magnetic field $B$, is nonzero, the junction supports both even-\(\omega\) intralead pairing
$F_{\rm LL/RR}=\langle \psi_{\rm L/R}\psi_{\rm L/R}\rangle$ and odd-$\omega$ interlead pairing $F_{\rm LR}=\langle \psi_{\rm L}\psi_{\rm R}\rangle$. Adapted from A.V. Balatsky \textit{et al}.~\cite{Balatsky2018}.}}
\label{fig:FigJosephson}
\end{figure}

\bl{Josephson and multi-terminal devices provide a second established exchange-sector route. In a Josephson junction, the left-right lead index can play the role of an effective orbital label in the $\dss\dpp\doo\dt$ classification. The same device can support conventional even-$\omega$ intralead correlations and interlead correlations whose parity under left-right exchange is controlled by tunneling, phase bias, voltage bias, and device geometry. A phase-biased junction therefore gives access to orbital-even and orbital-odd sectors, and in some cases to odd-$\omega$ interlead correlations~\cite{Balatsky2018,PhysRevB.92.024512}. The ring Josephson geometry shown in Fig.~\ref{fig:FigJosephson} provides a simple example. For weak tunneling between the two superconducting leads, the local interlead anomalous Green's function takes the form}
\begin{equation}\label{eq:Josephson_FLR}
\bl{
{\cal F}_{\rm LR}(\dr=0,i\omega_n)
=
\frac{
2\omega_n \pi^2 \rho^2 t
e^{i(\phi_{\rm L}+\phi_{\rm R})/2}
\sin(\phi_{\rm LR}/2)
}{
\omega_n^2+\Delta_0^2
}.
}
\end{equation}
\bl{The explicit proportionality to $\omega_n$ makes this interlead component odd in Matsubara frequency exactly at the interface, while the same device
also supports ordinary even-$\omega$ intralead correlations. Here $t$ is the tunneling amplitude, $\rho$ is the quasiparticle density of states, $\Delta_0$ is the superconducting gap magnitude in the leads, and $\phi_{\rm LR} = \phi_{\rm L} - \phi_{\rm R}$ is the phase difference. In a static equilibrium Josephson junction the center-of-mass phase can often be gauged away, but it is kept explicitly because in voltage-biased or otherwise time-dependent settings it can acquire nontrivial center-of-mass dynamics~\cite{Josephson1962,PhysRevB.80.205408}.}

\begin{figure}[t!]
\centering
\includegraphics[width=0.7\linewidth]{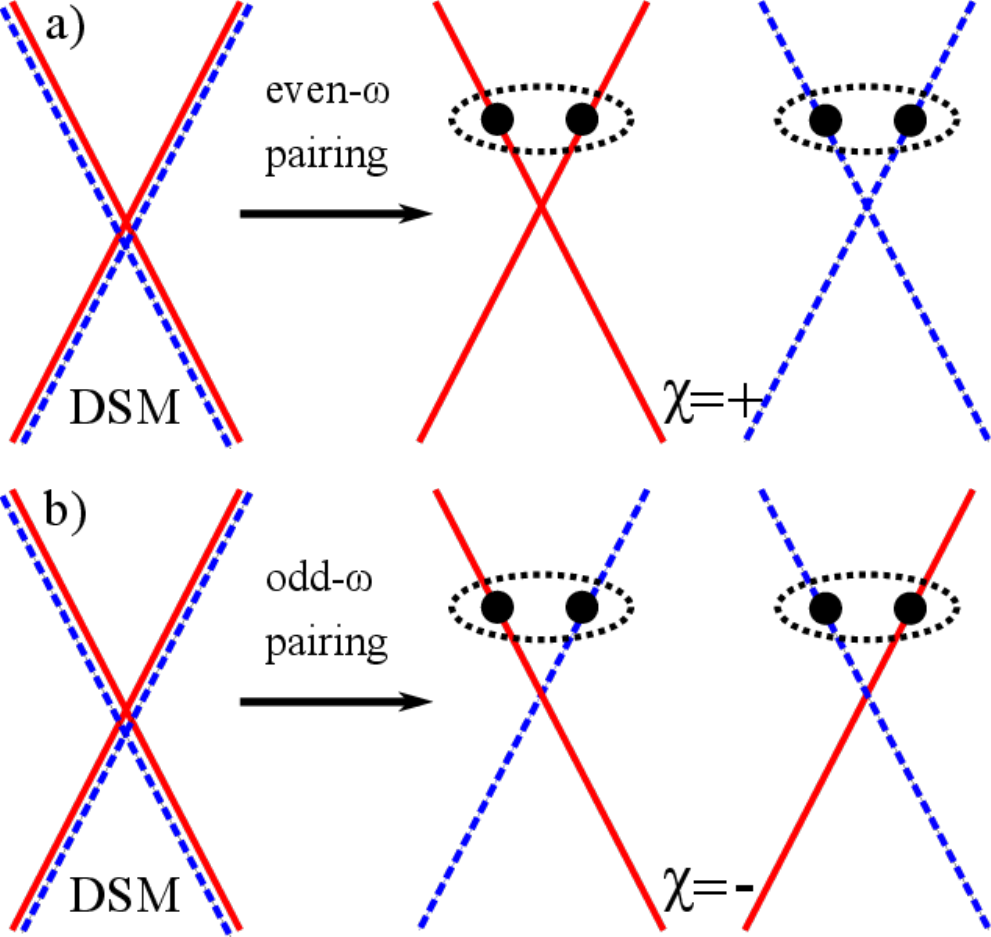}
\caption{\bl{Spin-singlet, $s$-wave pairing channels in Dirac semimetals for (a) even-$\omega$ and (b) odd-$\omega$ superconductivity. The even-$\omega$ gap corresponds to intra-chirality pairing, whereas the odd-$\omega$ Berezinskii channel is enabled by inter-chirality pairing. Red and blue dashed lines denote right- and left-handed quasiparticles, respectively. Adapted from P. O. Sukhachov \textit{et al}.~\cite{PhysRevB.100.180502}.}}
\label{fig:FigDirac}
\end{figure}

\subsubsection{Bulk-property dominated design}

\bl{Multiband, Dirac/Weyl, and spin-orbit-coupled superconductors provide a broader class of internal-index routes to unconventional exchange sectors. In multiband superconductors, the band index plays the role of an orbital label in the $\dss\dpp\doo\dt = -1$ classification. Band-symmetric anomalous correlations may remain conventional even-$\omega$, whereas band-antisymmetric correlations can become odd in frequency when the spin and spatial parities are otherwise conventional~\cite{PhysRevB.88.104514,PhysRevB.92.094517,Triola2020,PhysRevB.92.054516}. Thus, multiband systems naturally realize families of even- and odd-$\omega$ sectors within the same microscopic platform.}

\bl{Dirac and Weyl materials provide a closely related route, with chirality or helicity acting as an effective internal label. In the spin-singlet,
$s$-wave Dirac-semimetal model of Ref.~\cite{PhysRevB.100.180502}, the representative odd- and even-frequency gaps may be written as
\begin{subequations}
\begin{align}
\label{eq:Dirac_gaps.I}
\Delta_{\rm odd}(\omega) & = i\sigma_y\otimes\mathbb{I}_{2} \Delta_{\rm odd}(\omega), \\
\Delta_{\rm even}(\omega) & = \mathbb{I}_{2} \otimes \mathbb{I}_{2} \Delta_{\rm even}(\omega).
\end{align}
\end{subequations}
Here the first matrix acts in chirality space and the second in pseudospin space. The odd-$\omega$ spin-singlet $s$-wave state is antisymmetric in
chirality and therefore satisfies the constraint 
\begin{equation}\label{eq:Dirac_SPchiT}
\bl{
\dss \dpp \chi \dt = -1,
}
\end{equation}
the chirality analogue of the multiorbital $\dss \dpp \doo \dt=-1$ rule. Thus, Fig.~\ref{fig:FigDirac} illustrates a simple bulk mechanism: the even-$\omega$ channel pairs quasiparticles within the same chirality sector, whereas the odd-$\omega$ channel is enabled by inter-chirality pairing. Microscopically, the odd-$\omega$ gap is controlled by the frequency derivative of the pairing interaction, so that a repulsive but sufficiently frequency-dependent interaction can support Berezinskii pairing ~\cite{PhysRevB.100.180502}.

This Dirac-semimetal mechanism also gives a spectroscopic distinction: odd-$\omega$ pairing produces characteristic cusp-like structures in the density of states, unlike the more conventional spectral-weight redistribution associated with an even-$\omega$ gap~\cite{PhysRevB.100.180502}. Related internal-index routes may occur in Weyl-semimetal heterostructures, where spin-momentum locking and spin-polarized surface states promote unconventional even- and
odd-$\omega$ components~\cite{Parhizgar2020,PhysRevB.100.104511}.

Finally, in strongly spin-orbit-coupled superconductors, spin and orbital angular momenta are not separately conserved; the exchange classification
should instead be expressed in terms of pseudospin or total-angular-momentum sectors. This viewpoint is relevant for spin-$3/2$ systems, half-Heusler
materials, and related topological semimetals~\cite{PhysRevResearch.3.033255,PhysRevB.96.214514,PhysRevB.96.094526,Kim2018}.}

\subsection{Relative- and center-of-mass-space routes: \texorpdfstring{$\bm{\rho}$}{rho} and \texorpdfstring{$\drr$}{R} axes}

\bl{We next turn to routes along the spatial axes of the BA hypercube. The relative coordinate $\bm{\rho}$, or equivalently the relative momentum $\vk$, describes the internal orbital structure of the Cooper pair. Motion along this axis gives unconventional orbital pairing, such as $p$-, $d$-, or $f$-wave superconductivity. These states may be either even or odd in frequency depending on the remaining spin, orbital, and relative-time exchange parities.}

\bl{The center-of-mass coordinate $\drr$, or equivalently the center-of-mass momentum $\vQ$, describes spatial modulation of the condensate. The canonical example is the FFLO state, in which Cooper pairs condense at finite center-of-mass momentum. In the Fulde--Ferrell form the order parameter has a single plane-wave component,
\begin{equation}\label{eq:FF}
\Delta(\drr)=\Delta_{\vQ}e^{i\vQ\cdot\drr},
\end{equation}
while in the Larkin--Ovchinnikov form it is a standing-wave modulation, $\Delta(\drr) \sim \Delta_{\vQ}\cos(\vQ\cdot\drr)$~\cite{PhysRev.135.A550,Larkin1964}. PDW states generalize this idea to superconductors with intrinsic spatial modulation of the pair amplitude, often intertwined with charge or spin order~\cite{Daniel2020}. Nested Fermi surfaces provide a bulk route in which the same electronic structure selects both the finite pair momentum $\vQ$ and the relative-momentum structure of the Cooper pair~\cite{PhysRevLett.129.167001}. In the BA hypercube, these are Berezinskii Type $(+)$ nontrivial $\drr$-sector states.}

\bl{A more direct route to the mixed $(\tau,\drr)$ sector is obtained when finite-$\vQ$ superconductivity coexists with additional spatial order. For example, in a $d$-wave superconductor with charge-density-wave order, the charge modulation mixes momenta separated by the CDW wave vector and generates PDW components. In such a setting the anomalous Green's function contains both even- and odd-$\omega$ PDW components; schematically,
\begin{subequations}\label{eq:PDW}
\begin{align}
\label{eq:PDW.I}
&	\df(\vk,\vQ;i\omega_n) 				= \df_{\rm e}(\vk,\vQ;i\omega_n) + \df_{\rm o}(\vk,\vQ;i\omega_n), \\
\label{eq:PDW.II}
& 	\df_{\rm o}(\vk,\vQ;-i\omega_n) 	= - \df_{\rm o}(\vk,\vQ;i\omega_n).
\end{align}
\end{subequations}
This example illustrates a bulk spatial-modulation route to odd- and even-$\omega$ correlations. Unlike the interface or internal-index mechanisms of Sec.~\ref{sec.ex_sec}, the key ingredient here is finite center-of-mass momentum $\vQ$: PDW or FFLO order makes the anomalous Green's function depend simultaneously on $(\vk,i\omega_n)$ and $\vQ$. This spatial modulation can mix even- and odd-$\omega$ components, producing odd-$\omega$ PDW correlations alongside conventional even-$\omega$ ones~\cite{Chakraborty_2021}. Related work further emphasizes that finite-$\vQ$ pairing and finite-energy pairing are distinct, although their coexistence can also generate odd-$\omega$ components~\cite{Chakraborty_2022}.}

\subsection{Center-of-mass-time routes: \texorpdfstring{$T$}{T} or finite-\texorpdfstring{$\Omega$}{Omega} sectors}

\bl{The center-of-mass time coordinate \(T\), or equivalently the center-of-mass frequency $\Omega$, describes explicit temporal modulation of the superconducting state. This is distinct from odd-$\omega$ pairing: odd-$\omega$ refers to relative-time parity under $ \tau \rightarrow -\tau$, whereas finite-$\Omega$ refers to time dependence of the condensate as a whole. Thus, a time-modulated superconductor may be either even or odd in relative frequency.

A simple example is a voltage-biased Josephson junction, where the phase difference evolves as 
\begin{equation}\label{eq:JT}
\phi(t)=\phi_0+\Omega_J t, \qquad \Omega_J=\frac{2eV}{\hbar}.
\end{equation}
This produces superconducting correlations with nontrivial center-of-mass-time dependence, i.e. finite-$\Omega$ structure. Depending on other variables, this system can host both even- and odd-$\omega$ correlations~\cite{Josephson1962,PhysRevB.80.205408}.}

\bl{More generally, periodically driven or Floquet superconductors provide a direct route to the $T$-axis; experimentally, THz pump-probe experiments have demonstrated coherent condensate dynamics, including Higgs/amplitude oscillations~\cite{Matsunaga2014}. For a drive period $\mathcal{T}$, the anomalous correlation can be expanded as
\begin{equation}\label{eq:Floquet_F}
\df(T,\tau) = \sum_{m} e^{-im\Omega T}\df_m(\tau), \qquad \Omega=\frac{2\pi}{\mathcal{T}}. 
\end{equation}
The Floquet index $m$ acts as an additional effective label and can enable driven odd-$\omega$ components even when the corresponding static system is conventional~\cite{PhysRevB.103.104505}. In the BA-hypercube language, the drive activates the center-of-mass-time axis, while the relative-time parity is still determined by the behavior of $\df_m(\tau)$ under $ \tau \rightarrow -\tau$.}

\bl{The routes discussed above modify exchange labels, spatial structure, or center-of-mass time dependence. A different possibility is to modify the frequency structure of the anomalous Green's function itself, for example through open-system dynamics, non-Hermitian spectra, or self-conjugate Majorana degrees of freedom. These mechanisms should not be viewed as new symmetry classes; rather, they provide microscopic routes for generating strongly frequency-dependent anomalous correlations that may project onto even- or odd-$\omega$ sectors depending on the exchange structure, cf. Fig.~\ref{fig:Fig8}.}

\subsection{Unconventional routes: Majorana  systems}

\begin{figure}[t!]
\centering
\includegraphics[width=1.0\linewidth]{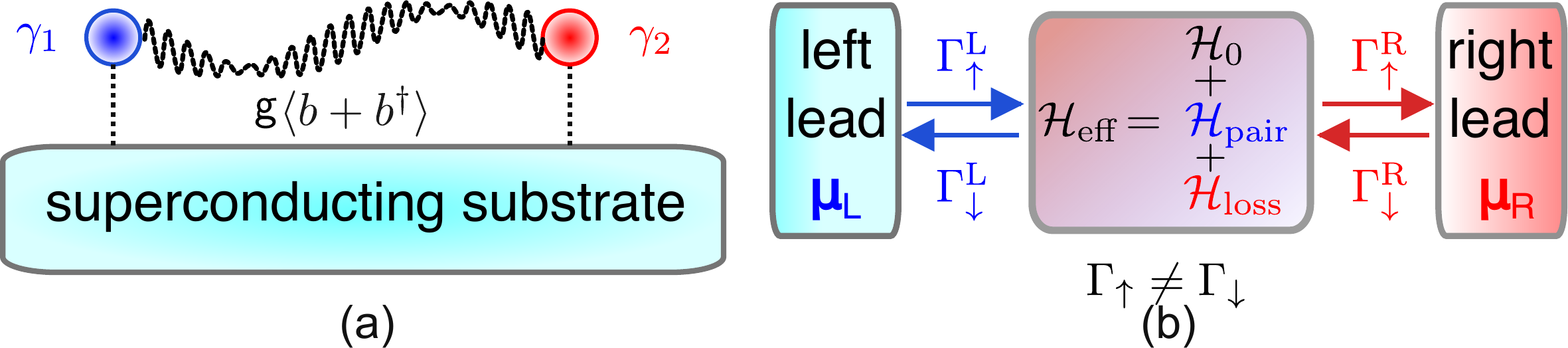}
\caption{\bl{Schematic frequency-structure routes beyond conventional Hermitian quasiparticles. (a): interacting Majorana modes (through some bosonic modes $b$) provide an intrinsic route to odd-$\omega$ correlations through their self-conjugate structure. (b): asymmetric dissipation in a paired system can generate odd-$\omega$ anomalous correlations, cf. Eq.~\eqref{eq:NH_BCS}.}}\label{fig:Fig8}
\end{figure}

\bl{Majorana excitations naturally demonstrate odd- $\omega$ pairing correlations. Indeed, the Majorana excitations  are ``particle" and ``hole" excitation at the same time: $\gamma_i = \gamma^{\dagger}_i$. The Green's function for Majorana state ${\cal G}(\tau_1,\tau_2) = \braket{T_{\tau} \gamma(\tau_1)\gamma(\tau_2)}$  is the ${\cal G}$ and $\df$ - anomalous pairing function at the same time. For the simplest example of the free Majorana fermion mode at zero energy we have ${\cal G}(i\omega_n) = \df(i\omega_n) = \frac{1}{i\omega_n}$
and the connection between the Majorana zero energy mode and the odd-$\omega$ pairing is obvious~\cite{PhysRevB.92.121404}.

For superconducting hetero-structures, the connection between Majorana modes in superconductor and odd-$\omega$ states was  emphasized by Tanaka and collaborators~\cite{Tanaka2012}. We give two examples of the odd-$\omega$ phase that naturally emerges in interacting Majorana models. \\

\noindent
a) The first one is the Sachdev-Ye-Kitaev (SYK) model used to describe the ``strange metal phases"~\cite{RevModPhys.94.035004}, where Majorana excitations interact via random quartic terms in an all-to-all interaction:
\begin{subequations}
\begin{align}
\label{eq.SYKHamiltonian.1}
{\cal H}_{\rm int} & = \sum_{ijkl = 1}^N \frac{J_{ijkl}}{4!} \gamma_i \gamma_j \gamma_k \gamma_l, \\
\label{eq.SYKHamiltonian.2}
E(J_{ijkl}) & = 0, \quad E(J^2_{ijkl}) = 6J^2/N.
\end{align}
\end{subequations}
Here $E(x)$ is expectation value of  random $x$ over distribution. The solution to this model is known in large $N$ limit and it gives the exact expression for Majorana propagator $G_{ij}(\tau_1,\tau_2) = \braket{ T_{\tau} \gamma_i(\tau_1)\gamma_j(\tau_2)}, \forall {i,j}$~\cite{Rosenhaus_2019}:
\begin{equation}\label{eq:SYKpropagator}
{\cal G}_{ij}(\tau_1,\tau_2) = \frac{\rm{sgn}(\tau_{12})}{(J\tau_{12})^2\Delta}, \;
\Delta = \frac{1}{4}, \;
\tau_{12} = \tau_1 - \tau_2.
\end{equation}
This solution demonstrates the stability of a single odd phase $\omega$ in the interacting fermion model as a single phase that is thermodynamically stable. This observation refutes the often claimed ``unstable nature"of the odd-$\omega$ state that can not exist on its own and always requires the conventional pairing state to ``feed off". \\

\noindent
b) The second example of odd-$\omega$ pairing of {\em interacting} Majorana fermions was the case of boson-mediated interaction between Majorana modes at the ends of topological wires. In that case authors claim the $\dss\dpp\doo\dt = -1$ applies to Majorana pairing states where the spatial swap now means the swap of the ends of the wire and swap of Majorana states on ends of wires keeps $S = 1$ for same spin states. Once one turns the   Majorana fermion- boson interaction, and it exceeds critical value ${\sf g} > {\sf g}_{\rm crit} = 1$, cf. Fig.~\ref{fig:Fig8}(a), the Majorana modes develop intra-wire odd-$\omega$  and cross wire even-$\omega$ correlations \cite{PhysRevB.92.121404}. 

We also comment that both examples  suggest that devices and platforms that host  Majorana states are the good medium to realize odd $\omega$ pairing states.}

\subsection{Unconventional routes: non-Hermitian systems}

\bl{A minimal non-Hermitian example is a paired fermionic mode with spin-dependent decay, cf. Fig.~\ref{fig:Fig8}(b),
\begin{equation}\label{eq:NH_BCS}
{\cal H}_{\rm eff} = \sum_{\vk\sigma} \left(\epsilon_{\vk}-i\Gamma_\sigma\right) c^\dagger_{\vk\sigma}c_{\vk\sigma} +
\sum_{\vk} \Delta c^\dagger_{\vk\uparrow}c^\dagger_{-\vk\downarrow} + {\rm h.c.}
\end{equation}
In the Nambu basis $\Psi_{\vk}=(c_{\vk\uparrow},c^\dagger_{-\vk\downarrow})^T$, the 
anomalous Green's function is ${\cal F}(\vk,\omega) = \Delta/D(\vk,\omega)$ where the denominator is 
\begin{equation}\label{eq:NH_F}
D(\vk,\omega) = \left(\omega-\epsilon_{\vk}+i\Gamma_\uparrow\right) \left(\omega+\epsilon_{\vk}-i\Gamma_\downarrow\right) -|\Delta|^2 .
\end{equation}
The odd-$\omega$ part, ${\cal F}_{\rm o}=[{\cal F}(\omega)-{\cal F}(-\omega)]/2$, is therefore proportional to the spin-asymmetric decay as
\begin{equation}\label{eq:NH_Fodd}
{\cal F}_{\rm o}(\vk,\omega) = - \frac{i(\Gamma_{\uparrow} - \Gamma_{\downarrow})\omega\,\Delta}{D(\vk,\omega)D(\vk,-\omega)}.
\end{equation}
Thus, dissipation alone does not generate pairing. The odd-$\omega$ component requires both a pairing source, \(\Delta\), and an internal lifetime asymmetry, \(\Gamma_\uparrow\neq\Gamma_\downarrow\). Exceptional points (in non-Hermitian systems) provide a particularly sharp setting where such non-Hermitian odd-frequency components can be enhanced~\cite{Bandyopadhyay2020,PhysRevB.105.094502}.}

\subsection{Weakly explored hybrid sectors of the BA hypercube}

\bl{The BA hypercube is most useful not only as a catalogue of known phases, but as a way to identify unexplored pair structures built from familiar ingredients. Some two-axis hybrids have already been discussed. For example, finite-$\vQ$ PDW correlations can acquire odd-$\omega$ structure in systems where superconductivity coexists with charge or spin modulation~\cite{Chakraborty_2021,Chakraborty_2022,PhysRevB.103.174512}. Similarly, periodic driving can generate odd-$\omega$ components in otherwise conventional superconductors by introducing Floquet sideband labels~\cite{PhysRevB.103.104505}. These examples provide useful anchors for the less explored sectors below.}

\bl{A natural next target is the driven PDW sector, where finite center-of-mass momentum and finite center-of-mass frequency are present
simultaneously. At the level of the anomalous correlation function this corresponds to components
\begin{equation}\label{eq:hybrid_R_T}
{\cal F}(\vk,\vQ;\omega_n,\Omega_m),
\qquad
\vQ\neq 0,\quad \Omega_m \neq 0 .
\end{equation}
Equivalently, the order may be represented schematically as
\begin{equation}\label{eq:driven_PDW}
\Delta(\drr,T)
=
\Delta_{\vQ,\Omega}
e^{i\vQ\cdot\drr-i\Omega T}
+
{\rm c.c.}
\end{equation}
This sector combines the PDW/FFLO direction of the hypercube with the time-modulated or Floquet direction. Candidate platforms include driven Josephson arrays, voltage-biased multi-terminal junctions, and materials with incipient PDW order subjected to periodic drive, in analogy with non-equilibrium PDW dynamics and driven competing-order settings~\cite{Malakhov2021,PhysRevB.90.024506,PhysRevB.93.195139}.}

\bl{A second  unexplored target is an odd-$\omega$ driven PDW sector, where the same finite-$\vQ$, finite-$\Omega$ structure also has odd relative-frequency parity,
\begin{equation}\label{eq:hybrid_tau_R_T}
{\cal F}_{\rm o}(\vk,\vQ;\omega_n,\Omega_m)
=
-
{\cal F}_{\rm o}(\vk,\vQ;-\omega_n,\Omega_m).
\end{equation}
This is the $(\tau,\drr,T)$ hybrid sector of the BA hypercube. It combines two known mechanisms: spatial modulation that can generate odd-$\omega$ PDW correlations in superconductors coexisting with charge or spin order~\cite{Chakraborty_2021,Chakraborty_2022}, while Floquet driving can generate odd-$\omega$ components through sideband structure~\cite{PhysRevB.103.104505}. A concrete route would therefore be to start from a PDW- or CDW-coexisting with superconductor and apply a periodic drive, voltage bias, or pump pulse, in analogy with non-equilibrium PDW dynamics and driven competing-order settings
~\cite{Malakhov2021,PhysRevB.90.024506,PhysRevB.93.195139}. Such a state could be probed by combining spatial probes of finite-$\vQ$ order, such as STM or quasiparticle interference, with time-resolved or ac Josephson probes sensitive to finite-$\Omega$ response~\cite{Daniel2020,Josephson1962,PhysRevB.80.205408,Matsunaga2014}.}

\subsection{A possible route to odd-$\omega$ spin-triplet PDW correlations \label{sec.odd-f-PDW}}

\bl{We now describe a scheme that motivates an odd-$\omega$ finite-momentum, or PDW-like, triplet component and its driven analogue. The construction is based on the spin-active superconductor--half-metal interface of Eschrig and L\"{o}fwander~\cite{Eschrig2008}, but with a spatially and temporally modulated tunneling matrix at the interface. The spatial modulation may arise from patterned interfaces, moir\'{e} structures, magnetic textures, strain, or a spatially modulated tunnel barrier. The temporal modulation may be produced by a microwave drive, gate modulation, optical pumping, or an ac voltage applied locally to the interface. The purpose of the construction is not to solve the full non-equilibrium proximity problem, but to show how the BA hypercube naturally suggests a route to a hybrid odd-$\omega$, finite-$\mathbf Q$, finite-$\Omega_D$ anomalous component in a half-metal.}

\bl{Consider a conventional spin-singlet BCS superconductor coupled to a half-metal through a spin-active interface. The anomalous Green's function of the parent superconductor is
\begin{subequations}
\begin{align}
\label{eq.oddf.1.1}
\df^{\rm SC}_{\alpha\beta}(\vk,i\omega_n)
& =
f(\vk,i\omega_n)(i\sigma_y)_{\alpha\beta}, \\
\label{eq.oddf.1.2}
f(\vk,i\omega_n)
& =
\frac{\Delta}{\omega_n^2+\xi_{\vk}^2+|\Delta|^2},
\end{align}
\end{subequations}
where $\xi_\vk=\epsilon_{\vk}-\mu$. In real space and imaginary time this parent anomalous function is even under relative-time exchange and even under relative-coordinate exchange for an $s$-wave superconductor. The half-metal is assumed to contain only the spin-$\uparrow$ band at low energies, with
\begin{equation}\label{eq.oddf.2}
G_{\uparrow}(\vk,i\omega_n)
=
\frac{1}{i\omega_n-\xi^{\rm HM}_\vk}.
\end{equation}
The tunneling Hamiltonian is
\begin{equation}\label{eq.oddf.3}
{\cal H}_T(t)
=
\sum_{\mathbf{r}}
d^\dagger_{\uparrow,\dr}
\left[
t_{\uparrow\uparrow}(\dr,t)c_{\uparrow,\dr}
+
t_{\uparrow\downarrow}(\dr,t)c_{\downarrow,\dr}
\right]
+{\rm h.c.},
\end{equation}
where $t_{\uparrow\uparrow}$ and $t_{\uparrow\downarrow}$ are, respectively, spin-conserving and spin-flip tunneling amplitudes. We allow both amplitudes to have weak spatial modulation along the plane of the interface and weak temporal modulation at drive frequency $\Omega_D$,
\begin{subequations}
\begin{align}
\label{eq.oddf.4.1}
t_{\uparrow\uparrow}(\dr,t)
& =
t_0+\delta t_0
\cos(\mathbf{Q}\cdot\dr)
\cos(\Omega_D t), \\
\label{eq.oddf.4.2}
t_{\uparrow\downarrow}(\dr,t)
& =
t_{\rm sf}+\delta t_{\rm sf}
\cos(\mathbf{Q}\cdot\dr+\phi)
\cos(\Omega_D t+\theta).
\end{align}
\end{subequations}
Here $\mathbf{Q}$ is the modulation wave vector along the interface, i.e. perpendicular to the tunneling direction, while $\phi$ and $\theta$ are the relative spatial and temporal phases between the spin-conserving and spin-flip modulations.} 

\begin{figure}[t!]
\centering
\includegraphics[width=1.0\linewidth]{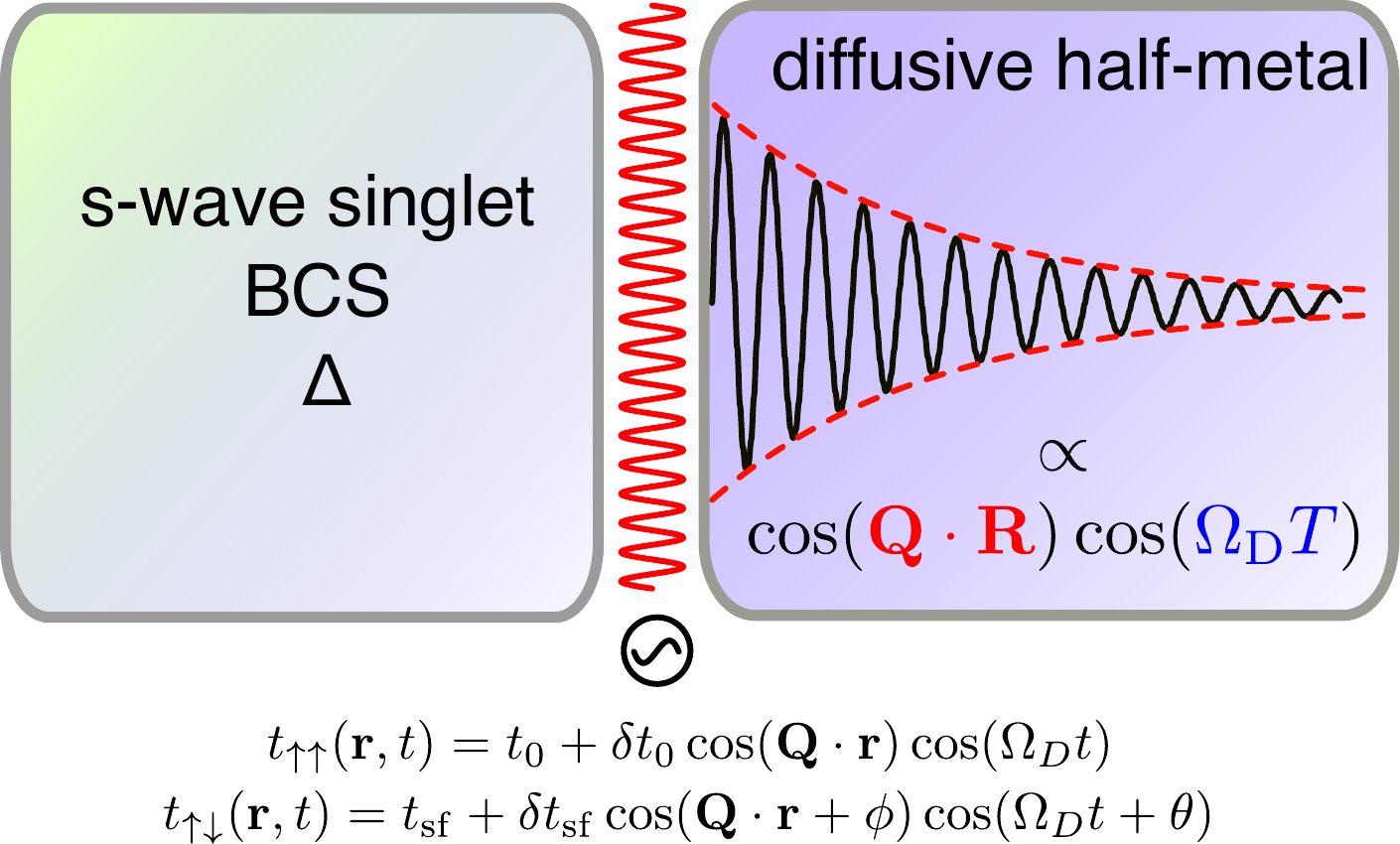}
\caption{\bl{Spin-active superconductor--half-metal interface realizing an odd-$\omega$, finite-$\mathbf Q$, finite-$\Omega_D$ anomalous response. The modulated interface tunneling generates the center-of-mass harmonics $\mathbf Q$ and $\Omega_D$, while the spin-active conversion produces an equal-spin triplet component in the diffusive half-metal. The spatial decay shown in the half-metal is schematic and follows from the Usadel description in the diffusive regime.}}\label{fig:Fig9}
\end{figure}

\bl{Integrating out the fermionic degrees of freedom on the superconducting side yields an induced equal-spin anomalous self-energy in the half-metal. In the time domain this object may be written schematically as
\begin{widetext}
\begin{equation}\label{eq.oddf.5}
\Sigma_{\uparrow\uparrow}(\dr_1,t_1;\dr_2,t_2)
=
f(\dr_1-\dr_2;t_1-t_2)
\left[
t_{\uparrow\uparrow}(\dr_1,t_1)t_{\uparrow\downarrow}(\dr_2,t_2)
-
t_{\uparrow\downarrow}(\dr_1,t_1)t_{\uparrow\uparrow}(\dr_2,t_2)
\right].
\end{equation}
\end{widetext}
The antisymmetric combination in Eq.~\eqref{eq.oddf.5} is the equal-spin analogue of the spin-active conversion process: it vanishes if the interface has no spin mixing or no spin-flip component. We now introduce the center-of-mass and relative variables ${\bf R} = \frac{\dr_1 + \dr_2}{2}$, $\bm{\rho}=\dr_1-\dr_2$, $T=\frac{t_1+t_2}{2}$, $\tau=t_1-t_2$. Keeping only terms linear in $\delta t_0$ and $\delta t_{\rm sf}$, Eq.~\eqref{eq.oddf.5} becomes
\begin{widetext}
\begin{equation}\label{eq.oddf.6}
\Sigma_{\uparrow\uparrow}(\mathbf{R},\bm{\rho};T,\tau)
=
f(\bm{\rho};\tau)
\left[
t_0\delta t_{\rm sf}\,
{\cal C}_{\phi,\theta}(\mathbf{R},\bm{\rho};T,\tau)
-
t_{\rm sf}\delta t_0\,
{\cal C}_{0,0}(\mathbf{R},\bm{\rho};T,\tau)
\right],
\end{equation}
\end{widetext}
where
\begin{widetext}
\begin{equation}\label{eq.oddf.7.1}
{\cal C}_{\phi,\theta}(\drr,\bm{\rho};T,\tau)
=
\cos\!\left[\mathbf{Q}\cdot\left(\drr-\frac{\bm{\rho}}{2}\right)+\phi\right]
\cos\!\left[\Omega_D\left(T-\frac{\tau}{2}\right)+\theta\right] -
\cos\!\left[\mathbf{Q}\cdot\left(\drr+\frac{\bm{\rho}}{2}\right)+\phi\right]
\cos\!\left[\Omega_D\left(T+\frac{\tau}{2}\right)+\theta\right].
\end{equation}
\end{widetext}
Using elementary trigonometric identities, the spin-flip modulation channel can also be written as
\begin{widetext}
\begin{align}
\label{eq.oddf.8}
{\cal C}_{\phi,\theta}(\drr,\bm{\rho};T,\tau)
& =
\sin\!\left(\mathbf{Q}\cdot\drr+\Omega_D T+\phi+\theta\right)
\sin\!\left(\frac{\mathbf{Q}\cdot\bm{\rho}+\Omega_D\tau}{2}\right) +
\sin\!\left(\mathbf{Q}\cdot\drr-\Omega_D T+\phi-\theta\right)
\sin\!\left(\frac{\mathbf{Q}\cdot\bm{\rho}-\Omega_D\tau}{2}\right).
\end{align}
\end{widetext}
This expression is exact in $\mathbf Q$ and $\Omega_D$ within the assumed single-harmonic interface modulation; no expansion in small $\mathbf Q$ is required. Eq.~\eqref{eq.oddf.8} also makes the symmetry content transparent. The induced source is modulated in the center-of-mass coordinate $\drr$ and center-of-mass time $T$, while its internal structure is odd under the simultaneous exchange $(\bm{\rho},\tau)\rightarrow(-\bm{\rho},-\tau)$, as required for an equal-spin anomalous amplitude. In the purely spatial case $\Omega_D=0$, the leading source is odd in $\bm{\rho}$ and describes a finite-$\mathbf Q$ triplet component. In the driven case $\Omega_D\neq 0$, even at $\bm{\rho}=0$ one obtains, for the spin-flip modulation channel,
\begin{widetext}
\begin{equation}\label{eq.oddf.9}
{\cal C}_{\phi,\theta}(\drr,0;T,\tau)
=
2\cos(\mathbf{Q}\cdot\drr+\phi)
\sin(\Omega_D T+\theta)
\sin\!\left(\frac{\Omega_D\tau}{2}\right),
\end{equation}
\end{widetext}
which is explicitly odd in the relative time $\tau$. Thus the temporal drive provides a direct route to an odd-$\omega$ equal-spin component, while the spatial modulation supplies the finite center-of-mass momentum.}

\bl{The same symmetry content can be summarized schematically in a mixed Wigner/Floquet representation. This representation is used only as a bookkeeping purpose: $\omega$ denotes the relative frequency, while $s\mathbf Q$ and $l\Omega_D$ label the center-of-mass spatial and temporal harmonics. For the tunneling model in Eqs.~\eqref{eq.oddf.4.1}--\eqref{eq.oddf.4.2}, the modulated spin-flip amplitude has the Fourier decomposition
\begin{widetext}
\begin{equation}\label{eq.oddf.10a}
\cos(\mathbf Q\cdot\drr+\phi)
\cos(\Omega_D t+\theta)
=
\sum_{s,l=\pm1}
\frac{e^{is\mathbf Q\cdot\drr}
e^{il\Omega_D t}
e^{i(s\phi+l\theta)}}{4}.
\end{equation}
\end{widetext}
The spin-conserving modulation has the same harmonics but with zero spatial and temporal phase,
\begin{equation}\label{eq.oddf.10b}
\cos(\mathbf Q\cdot\drr)
\cos(\Omega_D t)
=
\frac{1}{4}
\sum_{s,l=\pm1}
e^{is\mathbf Q\cdot\drr}
e^{il\Omega_D t}.
\end{equation}
Thus, for the harmonic proportional to
$e^{is\mathbf Q\cdot\drr}e^{il\Omega_D T}$, the induced equal-spin self-energy takes the schematic form
\begin{widetext}
\begin{equation}\label{eq.oddf.10}
\Sigma_{\uparrow\uparrow}^{(s,l)}(\vk;\omega)
\sim
\frac{1}{4}
\left[
t_0\delta t_{\rm sf}e^{i(s\phi+l\theta)}
-
t_{\rm sf}\delta t_0
\right]
\left[
f\!\left(\vk+\frac{s\mathbf Q}{2},
\omega-\frac{l\Omega_D}{2}\right)
-
f\!\left(\vk-\frac{s\mathbf Q}{2},
\omega+\frac{l\Omega_D}{2}\right)
\right],
\qquad s,l=\pm1 .
\end{equation}
\end{widetext}
The factor $e^{i(s\phi+l\theta)}$ therefore comes only from the spin-flip modulation in Eq.~\eqref{eq.oddf.4.2}; the spin-conserving modulation in Eq.~\eqref{eq.oddf.4.1} carries no additional phase. The two shifted arguments of $f$ arise from the two relative-coordinate and relative-time positions of the interface factors in Eq.~\eqref{eq.oddf.5}. Eq.~\eqref{eq.oddf.10} should therefore be read as the lowest-harmonic Wigner/Floquet representation of the anomalous source, not as a full non-equilibrium solution.}

\bl{With the same convention, the induced anomalous Green's function in the half-metal may be written schematically as
\begin{widetext}
\begin{equation}\label{eq.oddf.11}
\df_{\uparrow\uparrow}^{(s,l)}(\mathbf{k};\omega)
\sim
G_{\uparrow}
\!\left(\vk + \frac{s\mathbf Q}{2},
\omega+\frac{l\Omega_D}{2}\right)
\Sigma_{\uparrow\uparrow}^{(s,l)}(\mathbf{k};\omega)
\bar{G}_{\uparrow}
\!\left(\mathbf{k}-\frac{s\mathbf Q}{2},
\omega-\frac{l\Omega_D}{2}\right).
\end{equation}
\end{widetext}
Here $\bar{G}_{\uparrow}(\mathbf{k},\omega)$ is the corresponding hole propagator in the half-metal. Eq.~\eqref{eq.oddf.11} is again only a compact lowest-order expression. In a complete driven treatment, the Green's functions are matrices in Floquet or Keldysh-Floquet space, and Eq.~\eqref{eq.oddf.11} represents the weak-drive, single-harmonic component relevant for identifying the symmetry of the induced anomalous response. The odd-$\omega$ component is obtained by projection onto the part antisymmetric in the relative frequency,
\begin{equation}\label{eq.oddf.12}
\df_{\uparrow\uparrow}^{(s,l),{\rm odd}}(\vk;\omega)
=
\frac{1}{2}
\left[
\df_{\uparrow\uparrow}^{(s,l)}(\vk;\omega)
-
\df_{\uparrow\uparrow}^{(s,l)}(\vk;-\omega)
\right].
\end{equation}
This definition avoids any small-$\mathbf Q$ expansion and makes clear that the odd-$\omega$ character refers to relative-frequency parity, whereas $s\mathbf Q$ and $l\Omega_D$ label center-of-mass spatial and temporal modulation.}

\bl{The clean tunneling expression above should not be confused with a full solution of the diffusive proximity problem~\cite{Eschrig_2015}. In the quasi-classical description one separates the fast relative momentum from the slow center-of-mass variables by writing $\mathbf k\simeq k_{\rm F}\hat{\mathbf k}$ and integrating over the single-particle energy $\xi^{\rm HM}_{\mathbf k}$ measured from the Fermi level. The remaining dependence on $\hat{\mathbf k}$ labels angular harmonics of the relative motion. In a dirty half-metal, impurity scattering suppresses anisotropic angular harmonics, so the Usadel limit retains the isotropic component. This projection removes the relative-momentum dependence of the anomalous amplitude, but it does not remove the slow center-of-mass harmonics $s\mathbf Q$ and $l\Omega_D$. We denote the corresponding diffusive component by
\begin{equation}\label{eq.oddf.13}
\df_{\uparrow\uparrow}^{(s,l),{\rm qcl}}(\omega)
\equiv
\left\langle
\int d\xi^{\rm HM}_{\mathbf k}\,
\df_{\uparrow\uparrow}^{(s,l)}(\mathbf k;\omega)
\right\rangle_{\hat{\mathbf k}},
\end{equation}
where $\langle\cdots\rangle_{\hat{\mathbf k}}$ denotes angular averaging over the Fermi surface. If this projection yields a nonzero equal-spin triplet component, and the surviving isotropic component is even under $\dpp$, the constraint $\dss \dpp \doo \dt=-1$ therefore requires it to be odd in $\omega$,
\begin{equation}\label{eq.oddf.14}
\df_{\uparrow\uparrow}^{(s,l),{\rm qcl}}(-\omega)
=
-
\df_{\uparrow\uparrow}^{(s,l),{\rm qcl}}(\omega).
\end{equation}
Restoring the center-of-mass harmonics gives
\begin{equation}\label{eq.oddf.15}
\df_{\uparrow\uparrow}^{\rm qcl}(\mathbf R,T;\omega)
=
\sum_{s,l=\pm1}
\df_{\uparrow\uparrow}^{(s,l),{\rm qcl}}(\omega)
e^{is\mathbf Q\cdot\mathbf R - il\Omega_D T}.
\end{equation}
Eq.~\eqref{eq.oddf.15} represents the schematic isotropic equal-spin triplet component induced in the half-metal. Since each nonzero harmonic amplitude in this projected sector is odd in $\omega$, it describes an odd-$\omega$, finite-$\mathbf Q$, finite-$\Omega_D$ PDW-like anomalous component. A full calculation of its magnitude and boundary-condition dependence requires solving the corresponding time-dependent Usadel problem, and of course the heating issues due to the periodic drive. We do not attempt that calculation here. The purpose of the present construction is instead to identify the symmetry of the boundary-generated anomalous component. The clean tunneling analysis shows how the spatially and temporally modulated spin-active interface produces harmonics labeled by $s\mathbf Q$ and $l\Omega_D$, while the diffusive projection selects the isotropic equal-spin triplet sector in the half-metal. By the $\dss\dpp\doo\dt=-1$ constraint, any such isotropic equal-spin triplet component in a single-orbital half-metal must be odd in $\omega$, provided the averaging produced $\dpp = 1$ uniform component. Thus the interface provides a symmetry-allowed route to an odd-$\omega$, finite-$\mathbf Q$, finite-$\Omega_D$ PDW-like anomalous response. Its magnitude, decay length, and stability are separate microscopic questions that require a full quasi-classical non-equilibrium treatment. The schematic form of the odd-$\omega$ PDW correlation function is shown in Fig.~\ref{fig:Fig9}.}

\section{Stability constraints and experimental diagnostics \label{sec:stability}}

\bl{The BA hypercube should be read as a symmetry roadmap, not as a guarantee of thermodynamic stability. A sector allowed by the exchange constraint $\dss\dpp\doo\dt=-1$ may appear as an induced anomalous correlation, a transient driven response, or a stable bulk phase. These are physically different situations. A stable ordered state must also have positive stiffness against phase, spatial, and temporal fluctuations.}

\bl{For finite-$\vQ$ sectors, such as FFLO or PDW order, the simplest long-wavelength Landau free energy contains a stiffness cost for spatial modulation~\cite{Daniel2020,RevModPhys.77.935},
\begin{equation}\label{eq:stability_GL_Q}
{\cal L}_{\vQ} 
=
\left[
r
+
K\left(\vQ-\vQ_0\right)^2
\right]
|\Delta_{\vQ}|^2
+
u|\Delta_{\vQ}|^4
+\cdots .
\end{equation}
A finite-$\vQ$ condensate is stable only when the quadratic coefficient is negative near some nonzero $\vQ_0$, while the stiffness and quartic terms keep the state bounded. Disorder is especially restrictive because elastic scattering destroys the phase coherence required to maintain a finite center-of-mass momentum. This is why conventional FFLO and many PDW states favor clean or spatially organized systems. Hybrid odd-$\omega$ finite-$\vQ$ sectors are expected to be even more constrained, because they must maintain both spatial coherence and a well-defined relative-time structure~\cite{PhysRevB.101.094506}.}

\bl{For odd-$\omega$ sectors, the stability question is not determined by frequency parity alone. In a functional-integral formulation with a retarded pairing kernel, the quadratic action for a frequency-dependent gap can be written schematically as~\cite{PhysRevB.79.132502}
\begin{widetext}
\begin{equation}\label{eq:stability_oddw_kernel}
{\cal S}^{(2)}
=
T
\sum_{\omega_n,\omega_m}
\Delta^*(i\omega_n)
{\cal K}(i\omega_n,i\omega_m)
\Delta(i\omega_m),
\qquad
\Delta(-i\omega_n)=-\Delta(i\omega_n).
\end{equation}
\end{widetext}
In a basis that diagonalizes the pairing kernel, this reduces to the more familiar schematic form
\begin{equation}\label{eq:stblity}
{\cal S}^{(2)}
=
T
\sum_{\omega_n}
\Delta^*(i\omega_n)
\left[
D^{-1}(i\omega_n)-\Pi(i\omega_n)
\right]
\Delta(i\omega_n),
\end{equation}
where $D$ and $\Pi$ are the bare interaction Kernel and Cooper pair susceptibility, respectively. A stable odd-$\omega$ phase requires an instability of the kernel in the odd-frequency channel while preserving a physical, positive electromagnetic response. This is a stronger condition than merely finding an odd-$\omega$ component in the anomalous Green's function. Induced odd-$\omega$ amplitudes in proximity structures can be robust, especially when they are $s$-wave in the relative coordinate, but establishing an intrinsic homogeneous odd-$\omega$ bulk phase remains more subtle~\cite{PhysRevB.79.132502,PhysRevB.91.144514}.}

\bl{Driven sectors introduce a separate constraint: the drive can populate finite-$\Omega$ or Floquet sideband components, but heating and relaxation compete with superconducting coherence. A useful phenomenological criterion is that the induced hybrid component must live longer than the relevant dephasing time,
\begin{equation}\label{eq:stability_drive}
\Gamma_{\rm heat}+\Gamma_{\rm qp}
\ll
\Delta_{\rm sc},\Omega ,
\end{equation}
where $\Gamma_{\rm heat}$ is the drive-induced heating rate, $\Gamma_{\rm qp}$ is the quasiparticle relaxation rate, and $\Delta_{\rm sc}$ is the superconducting energy scale. High-frequency or prethermal Floquet regimes, voltage-biased junctions with controlled dissipation, and pump-probe settings with short observation windows are therefore the most natural platforms for finite-$\Omega$ hybrid sectors~\cite{Sacha2018,Else2020,Matsunaga2014}.}

\bl{The diagnostic strategy should follow the active axes of the BA hypercube. Spatial modulation along the $\drr$-axis can be probed by STM, quasiparticle interference, NMR line shapes, or diffraction-sensitive probes. Internal orbital structure along the $\bm{\rho}$-axis can be tested by nodal spectroscopy and phase-sensitive Josephson measurements. Phase-sensitive Josephson measurements, surface Andreev bound states, and spin-orbit-coupled Josephson platforms provide complementary probes of unconventional pairing symmetry and its higher-angular-momentum or topological descendants~\cite{ZuticMazin2005,Sengupta2001,Alidoust2021cSOC,Amundsen2024RMP}. Relative-time structure along the $\tau$-axis requires probes sensitive to frequency dependence, such as tunneling spectroscopy, Josephson response, and Meissner or spin-susceptibility measurements. Center-of-mass-time structure along the $T$-axis requires pump-probe spectroscopy, ac Josephson response, Floquet sidebands, or time-resolved ARPES. Controlled disorder studies are particularly important for distinguishing robust induced correlations from fragile bulk ordered phases.}

\bl{Thus, the practical use of the BA hypercube is twofold. First, it identifies which pairing structures are allowed by exchange symmetry. Second, it makes clear which additional stability filters must be passed before a symmetry-allowed sector can be regarded as a realizable superconducting phase.}

\section{Future directions and summary}

\bl{We have presented a symmetry-based classification of superconducting pairing orders formulated at the level of the anomalous Green's function. The
classification follows from the exchange parities of spin, relative space, orbital or internal labels, and relative time, constrained by $\dss\dpp\doo\dt=-1$. This reproduces the familiar distinction between spin-singlet and spin-triplet pairing and between $s$-, $p$-, and $d$-wave orbital channels, while placing even- and odd-frequency pairing on the same footing.}

\bl{By separating relative and center-of-mass space-time variables, we organized the allowed pair amplitudes into the BA hypercube. In
this representation, $\bm{\rho}$ describes the internal orbital structure of the Cooper pair, $\tau$ its relative-time or frequency parity, $\drr$
finite-momentum or spatially modulated order such as FFLO and PDW states, and $T$ explicit temporal modulation of the condensate. This construction
connects conventional BCS, unconventional orbital pairing, odd-$\omega$ pairing, FFLO/PDW order, and driven or Floquet superconducting correlations
within one framework.}

\bl{The   value of the BA hypercube  representation is that it identifies hybrid sectors that are symmetry allowed but not yet systematically explored. Examples include
odd-$\omega$ FFLO or PDW correlations, driven PDW states, and odd-$\omega$ driven finite-$\vQ$ superconducting correlations. Guided by the BA classification we developed the microscopic scheme where one can induce odd f PDW state. These states should be viewed as targets suggested by symmetry. The physical realization  of proposed new states requires additional microscopic analysis, including sufficient phase stiffness, controlled disorder, favorable pairing kernels, and, in driven systems, manageable heating and relaxation.}

\bl{The same swap logic can also be formulated for non-equilibrium system using contour-ordered Green's functions on the Schwinger--Keldysh contour. In that setting, the relative-time swap remains distinct from physical time reversal, while explicit center-of-mass-time dependence corresponds to finite-$\Omega$ or Floquet sectors. 
Throughout this work we have assumed ordinary fermionic exchange, so that the relevant pairwise swap operations have real $\mathbb{Z}_2$ eigenvalues.

}

\section{Acknowledgments \label{sec:sec.5}}

We are grateful to Igor \v{Z}uti\'c, Yukio Tanaka, Annica M. Black-Schaffer, Matthias Eschrig, Holger Fehske, and other numerous colleagues and collaborators for the discussions on the origin and stability of unconventional superconductivity over the years. Their arguments and critiques helped to shape our thinking. This work was supported by the U.S. Department of Energy, Office of Science, and Office of Basic Energy Sciences under award number DE-SC-0025580 (concept, writing) and  by the European Research Council under the European Union Seventh Framework ERS-2018-SYG 810451 HERO. SB acknowledges support from the University of Greifswald, Germany and the Office of Basic Energy Sciences, Material Sciences and Engineering Division, U.S. DOE, under Contract No. DE-FG02-99ER45790. 

\section{Author contributions}

A.B. discovered the initial classification and the organization of superconducting orders as Berezinski-Abrahams hypercude.  A.B and S.B. further developed the idea into the Berezinskii--Abrahams hypercube framework, developed model for the odd-$\omega$ PDW and led the analysis and manuscript preparation. Both authors discussed the results and approved the final version.



\appendix

\bl{\section{Swap properties in Schwinger-Keldysh contour \label{sec.app.1}}}

\begin{figure}[b!]
\centering 
\includegraphics[width=1.0\linewidth]{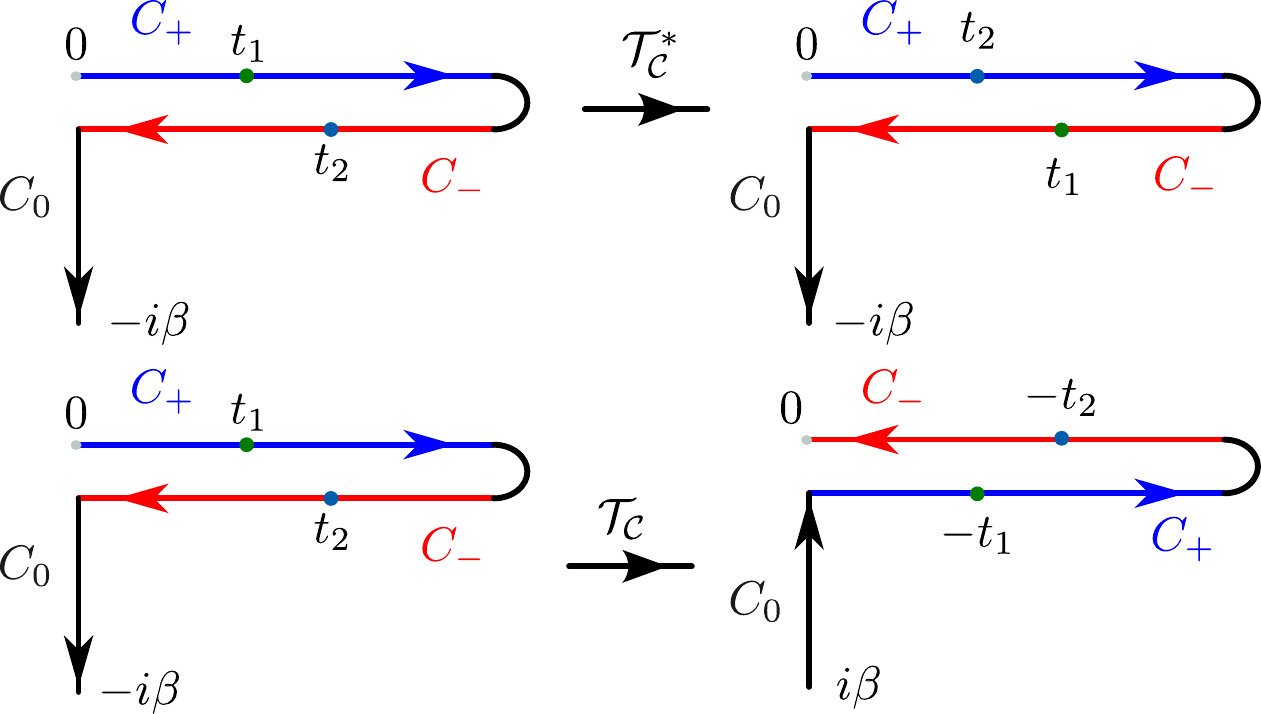}
\caption{\bl{Contour-based illustration of the time-swap operation $\dt_{\cal C}$ and the physical time-reversal operation ${\cal T}_{\cal C}$ on the Schwinger--Keldysh contour. Top: $\dt_{\cal C}$ exchanges the two contour-time arguments $t_1$ and $t_2$, corresponding to the non-equilibrium generalization of relative-time permutation. Bottom: ${\cal T}_{\cal C}$ reverses the physical time coordinates, $ t_i \rightarrow -t_i$, and acts as the usual time-reversal operation. This figure is the contour analogue of the real-time illustration shown in Fig.~\ref{fig:Fig1}.}}\label{fig:AFig1}
\end{figure}

\bl{In this Appendix, we describe how the exchange/swap operations introduced in the main text are generalized to a non-equilibrium setting with explicit center-of-mass time dependence. In this case, the anomalous Green's function is naturally defined on the Schwinger--Keldysh contour \({\cal C}\) as
\begin{equation}\label{aeq.1}
{\cal F}_{\alpha\beta,ab}(1,2)
=
\left\langle
{\cal T}_{\cal C}
c_{\alpha a}(1)c_{\beta b}(2)
\right\rangle,
\end{equation} \\
where $1=(\dr_1,z_1)$, $2=(\dr_2,z_2)$, $(z_1,z_2) \in {\cal C}$, and ${\cal T}_{\cal C}$ denotes contour ordering. The contour consists of the forward branch $C_+$, the backward branch $C_-$, and, when needed, the imaginary-time branch used to encode the initial density matrix.

The time-swap operation $\dt_{\cal C}$ acts by exchanging the two contour-time arguments,
\begin{equation}\label{aeq.2}
\dt_{\cal C}:\quad z_1 \leftrightarrow z_2,
\end{equation}
while leaving the physical meaning of the contour branches unchanged. This operation is the contour analogue of the relative-time permutation discussed in the main text. By contrast, the physical time-reversal operation ${\cal T}_{\cal C}$ reverses the physical time coordinates and acts antiunitarily on the microscopic fermionic operators. Thus, $\dt_{\cal C}$ is an exchange operation on the two-particle correlator, whereas ${\cal T}_{\cal C}$ is a physical symmetry operation.

The distinction is illustrated in Fig.~\ref{fig:AFig1}. In the upper panel, $\dt_{\cal C}$ interchanges the two contour-time positions $t_1$ and $t_2$ on the Keldysh contour. In the lower panel, ${\cal T}$ reverses the physical time coordinates, mapping $t_i \rightarrow -t_i$, and correspondingly reverses the orientation of the contour representation. The spatial operations are generalized analogously: $\dpp$ exchanges the two spatial coordinates $\dr_1 \leftrightarrow \dr_2$, while the physical inversion operation ${\cal I}$ maps each coordinate as $\dr_i \rightarrow -\dr_i$. Therefore, the exchange-based classification $\dss \dpp \doo \dt=-1$ can be formulated for contour-ordered anomalous Green's functions in the same way as for equilibrium time-ordered correlators.}

\clearpage
\pagebreak
\bibliography{References}

\end{document}